\documentclass[]{elsart}
\usepackage{graphicx}

\usepackage{amssymb}

\usepackage{natbib}

\begin{document}

\begin{frontmatter}



\title{Anisotropic magnetization, specific heat and resistivity of
$R$Fe$_2$Ge$_2$ single crystals}


\author{M. A. Avila\corauthref{cor1}}
\corauth[cor1]{Corresponding author}
\ead{avila@ameslab.gov},
\author{S. L. Bud'ko},
\author{P. C. Canfield}%
\address{Ames Laboratory and Department of Physics and Astronomy\\
Iowa State University, Ames, IA 50011}

\begin{abstract}
We have grown $R$Fe$_2$Ge$_2$ single crystals for $R$ = Y and ten
members of the lanthanide series (Pr, Nd, Sm, Gd-Tm, Lu) using Sn
flux as the solvent. The method yields clean, high quality crystal
plates as evidenced by residual resistivities and \textit{RRR}
values in the range of 3-12 $\mu\Omega$ cm and 20-90 respectively.
The crystals are also virtually free of magnetic impurities or
secondary phases, allowing the study of the intrinsic anisotropic
magnetic behavior of each compound. Characterization was made with
X-Ray diffraction, and temperature and field dependent
magnetization, specific heat and resistivity. Very strong
anisotropies arising mostly from CEF effects were observed for all
magnetic rare earths except Gd. Antiferromagnetic ordering
occurred at temperatures between 16.5 K (Nd) and 1.1 K (Ho) that
roughly scale with the de Gennes factor for the heavy rare earths.
For some members there is also a lower temperature transition
associated with changes in the magnetic structure. Tm did not
order down to 0.4~K, and appears to ba a van Vleck paramagnet. All
members which ordered above 2~K showed a metamagnetic transition
at 2~K for fields below 70~kOe. The calculated effective moments
per rare earth atom are close to the expected free ion values of
$R^{3+}$ except for Sm which displays anomalous behavior in the
paramagnetic state. The non-magnetic members of this series (Y,
Lu) are characterized by an unusually large electronic specific
heat coefficient ($\gamma\sim60$~mJ/mol~K$^2$) and
temperature-independent susceptibility term ($\chi_0\sim0.003$
emu/mol), indicative of a relatively large density of states at
the Fermi surface.
\end{abstract}

\end{frontmatter}

\maketitle

\section{Introduction}

The $R$T$_2$X$_2$ family of ternary intermetallic compounds
($R$~=~Y, La-Lu; T~=~Mn-Cu, Ru, Rh, etc and X~=~Si, Ge) have been
intensively studied for several decades due to the wide range of
physical behaviors displayed by its members\cite{szyt94a}. Most
$R$T$_2$X$_2$ compounds form in the ThCr$_2$Si$_2$ structure,
space group $I4/mmm$. This body-centered structure has a single
$R$ site in tetragonal point symmetry which can give rise to
highly anisotropic local moments at low temperatures, influencing
the magnetic behavior in a tractable way that allows many of these
compounds to be used as model systems.

Some works on the $R$Fe$_2$Ge$_2$ series of this family have
successfully established their chemical behavior such as crystal
structure, lattice parameters and chemical
bonds\cite{ross78a,bara90a,vent96a} and melting
temperatures\cite{moro97a,moro98a}. But so far the ground state
properties of the $R$Fe$_2$Ge$_2$ series haven't been as
intensively studied as others (e.g. $R$Cu$_2$Si$_2$ or
$R$Ni$_2$Ge$_2$), in part due to the fact that, whereas in these
compounds Fe is in its non-magnetic spin-paired $3d^6$
state\cite{bara90a}, many of the Fe-based impurities are strongly
magnetic and, even in small quantities, can prevent precise
determination of the main compound's bulk physical properties. For
example, an early report claiming the occurrence of partial
spontaneous magnetization of the Fe sub-lattice in some arc-melted
$R$Fe$_2$Ge$_2$ ingots\cite{feln75a} was not confirmed in
subsequent studies (nor in this present one) and was almost
certainly due to the presence of ferro- or ferri-magnetic second
phases in the samples.

The understanding of the anisotropic magnetic behavior and
ordering temperatures in this series has also suffered from the
lack of single crystals of sufficient size and quality to allow
direct measurements of orientation-dependent properties. Although
small single crystals picked out of annealed ingots have been used
for x-ray refinements\cite{vent96a}, larger crystals have been
reported only for LaFe$_2$Ge$_2$ and
CeFe$_2$Ge$_2$\cite{ebih95a,suga99a,suga00a} grown by the
Czochralski method. The latter compound has received special
attention\cite{ebih95a,suga99a,suga00a,neif85a,ansa88a,budk99a}
for being a non-magnetic heavy fermion system with
$\gamma=210$~mJ/mol K$^2$.

In this work, we present a detailed characterization of the
anisotropic ground state properties for 11 compounds of the
$R$Fe$_2$Ge$_2$ series ($R$ = Y, Pr, Nd, Sm, Gd-Tm, Lu) grown as
clean single crystal plates (with dimensions as large as
$4\times4\times0.2$~mm$^3$) by the flux growth method using Sn as
the solvent\cite{fis89a,can92a,can01a}, a method which also helps
avoid the inclusion of magnetic second phases in the samples.
After describing the experimental procedures used for crystal
growth and characterization, results will be presented separately
for each compound including a discussion of whether our
experiments confirm, correct or contradict any established or
claimed properties reported by previous works, such as ordering
temperatures and effective moments. We will finish with a
discussion on the trends observed along the series such as the
dependence of ordering temperatures and effective/saturated
moments on the de Gennes factor $(g_J-1)^2J(J+1)$.

\section{Experimental Details}

All single crystals of the $R$Fe$_2$Ge$_2$ series studied here
were grown out of Sn flux\cite{fis89a,can92a,can01a}. The crystals
grow quite easily and for a relatively wide range of growth
parameters. A typical procedure involved adding to a 2~ml alumina
crucible about 5~g of Sn (99.99\% purity), and 3-10 \textit{at}\%
of (at least 99.95\% purity) $R$, Fe and Ge elements in or close
to the ratio of 1:2:2 respectively. A small excess of Fe seemed to
facilitate the growths, and in several cases a ratio of 1:2.4:2.0
was used, although it was not crucial nor did it change the actual
measured properties of the crystals. The crucible with starting
elements was sealed in a quartz ampoule under partial argon
atmosphere, which was then placed in a box furnace. The elements
were dissolved in Sn by holding the temperature at 1200$^{\circ}$C
for 1-2 hours, then the crystals grew while the temperature was
reduced over 3-6 days to a chosen temperature varying between
500$^{\circ}$C and 800$^{\circ}$C (depending on solute
concentration), at which point the ampoule was quickly removed
from the furnace and the molten Sn flux decanted.

\begin{figure}[thb]
\includegraphics[angle=0,width=33pc]{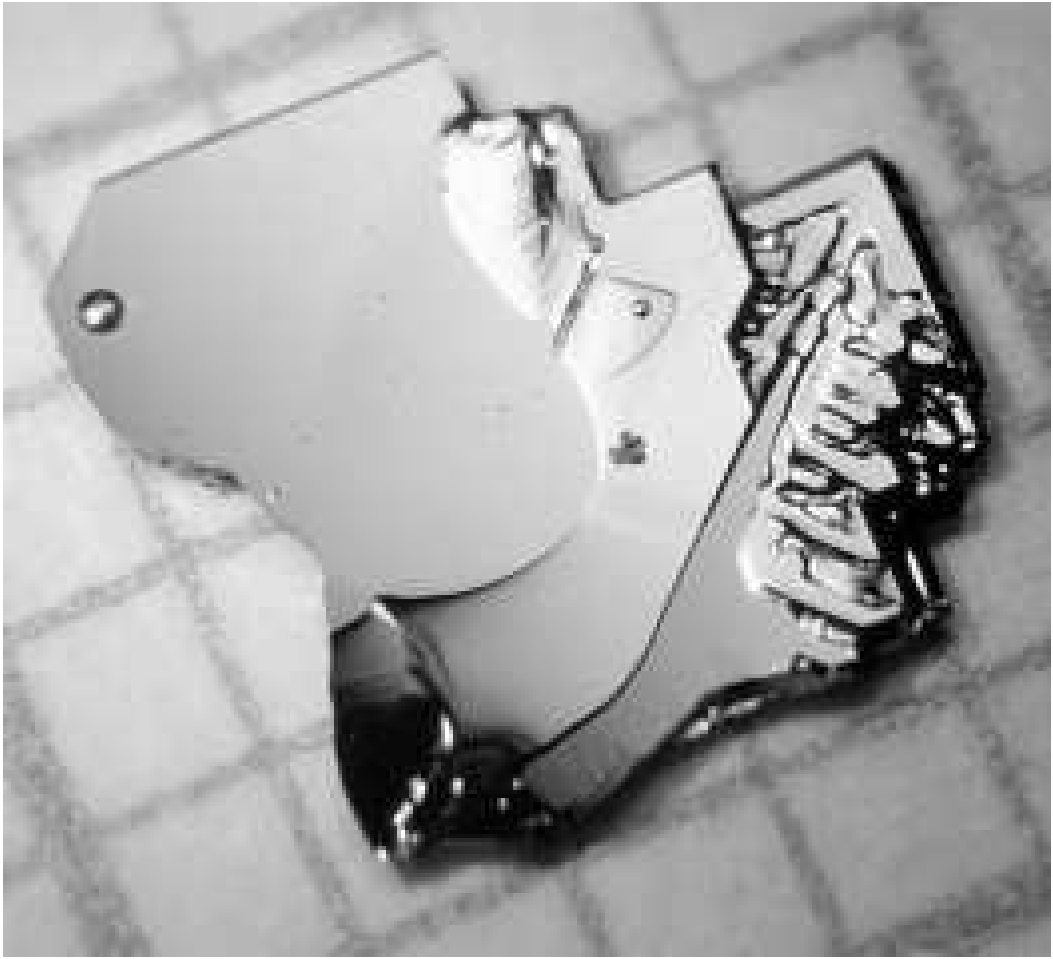}
\caption{\label{fcrystal} As-grown single crystal of
TbFe$_2$Ge$_2$ on a millimeter-scale paper. The round droplet on
the upper left is solidified Sn flux. Dendritic growth along the
[110] directions can be seen on the right side of the crystal.}
\end{figure}

The crystals form with a plate-like morphology as shown in
fig.~\ref{fcrystal}, with the \textit{c}-axis normal to the plane
of the plate. They typically have very smooth and clean surfaces,
and any remaining Sn flux droplets solidified on a surface (see
upper left of fig.~\ref{fcrystal}) can be easily removed with a
scalpel or polish. Most crystals have at least one very well
defined facet along the [100] \textit{a}-axis. The incomplete
surface formation seen on the right part of the crystal in
fig.~\ref{fcrystal} gives an insight into the growth dynamics,
with more rapid dendritic growth in the [110] directions.

DC magnetization measurements as a function of field (up to 55 kOe
or 70 kOe) and temperature (1.8 to 350 K) were performed in
Quantum Design SQUID magnetometers. Temperature sweeps were
performed on warming, after zero-field cooling the sample to 1.8~K
and applying a measuring field of 1~kOe. The samples were manually
aligned to measure the magnetization along the appropriate axis.
The notation $H||c$ or $\chi_c$ denotes measurements made with the
applied field along the $c$-axis (field perpendicular to the plate
shown in fig.~\ref{fcrystal}). $H\perp{c}$ or $\chi_{ab}$ denotes
measurements made with the field in the basal plane (along the
plate shown in fig.~\ref{fcrystal}), and $H||a$ or $\chi_a$
denotes measurements made with the field along a specific in-plane
orientation: $H||[100]$. Polycrystalline averages were calculated
by $\chi(T)=[2\chi_{ab}(T)+\chi_{c}(T)]/3$. These averages were
used to obtain the high temperature effective moments of the
magnetic rare earths, using a Curie-Weiss law:
$\chi(T)=C/(T-\theta_p)+\chi_0$, including a temperature
independent term to account for the relatively high susceptibility
found in the non-magnetic members of the series. Values of
$\chi_0$ used are presented along with other measured properties
in table~\ref{tbl1}. Ne\'{e}l temperatures were determined by the
maxima in $d(\chi T)/dT$, the temperature dependence of which near
the transition of an antiferromagnet is similar to that of the
magnetic specific heat\cite{fis62a}. This derivation procedure
enhances noise and two spurious features frequently appear in
these data: one just above 4.2~K due to difficulties in the MPMS
temperature control when the systems pass through the He boiling
point, and another around 12~K most likely due to the fact that
the MPMS systems change thermometers in this region.

Heat Capacity measurements were made on Quantum Design PPMS
systems and in some cases a $^3$He cooling option was installed,
allowing measurements down to $\sim0.4$~K. Before each run the
sample holder + grease background was measured for later
subtraction from the sample + background data. Estimate of the
magnetic specific heat of samples with moment-carrying rare earths
was made rather difficult due to the non-trivial behavior of the
measured specific heats $C_p(T)$ even for $R=$~Y and Lu (shown in
the following section). Similar difficulties have been reported
for polycrystalline PrFe$_2$Ge$_2$\cite{mala93a}. To obtain our
best estimate of the magnetic contribution
$C_m^{(R)}(T)=C_p^{(R)}(T)-C_{nm}^{(R)}(T)$, where $C_p^{(R)}(T)$
is the measured specific heat and $C_{nm}^{(R)}(T)$ is the
non-magnetic contribution for the rare-earth $R$, we fitted the
specific heats of LuFe$_2$Ge$_2$ and YFe$_2$Ge$_2$ in the region
11~K~$<T<$~45~K with a polynomial $\gamma T+AT^3+BT^5+CT^7$. The
region below 11~K was avoided in the fits because of some low
temperature features appearing in both samples which will be
discussed later, but the resulting polynomials were then extended
to represent $C_p^{(Lu)}(T)$ and $C_p^{(Y)}(T)$ in the full
temperature range. We then used an interpolation formula
$C_{nm}^{(R)}(T)=C_p^{(Lu)}(T)-(C_p^{(Lu)}(T)-C_p^{(Y)}(T))[(M_{Lu}^{3/2}-M_{R}^{3/2})/(M_{Lu}^{3/2}-M_{Y}^{3/2})]$,
where $M$ is the atomic mass of the respective rare earth
indicated in the subscript, to remove the non-magnetic part
$C_{nm}^{(R)}(T)$ of each compound's specific heat. This procedure
gives a better estimate of the non-magnetic contribution than
direct subtraction of LuFe$_2$Ge$_2$ and avoids the unphysical
situation of $C_p^{(R)}(T)<C_{nm}^{(R)}(T)$ for a wider
temperature range.

The magnetic entropy $S_m(T)$ of each sample was estimated by
numerical integration of $C_m(T)/T~vs.~T$. A simple linear
extrapolation from the data measured at the lowest temperature
down to the origin was used to represent $C_m(T)$ in the
unmeasured region, in general the errors in $S_m(T)$ introduced by
this simplification are smaller than those resulting from the
subtraction of the non-magnetic contributions discussed above. To
give an idea of the latter, in figure~\ref{fNd1}b we have included
an estimation of $S_m(T)$ resulting after direct subtraction of
$C_p(T)$ for LuFe$_2$Ge$_2$ to obtain $C_m(T)$ for NdFe$_2$Ge$_2$.
At 20~K the difference adds up to only 4\%. $S_m(T)$ is useful to
give a basic idea of the degeneracy $Z$ of the ground states
involved in the magnetic ordering process, since at the transition
$S_m(T)$ should approach R$lnZ$, where $R=8314$~mJ/mol~K is the
universal gas constant. For some compounds the interference of
nuclear Schottky anomalies at low temperatures as well as closely
spaced, higher temperature ($T\gtrsim T_N$) Schottky anomalies
from low-lying CEF levels did not allow a reliable estimate of
magnetic entropies.

Electrical resistance measurements were performed on Quantum
Design PPMS systems or on Quantum Design MPMS systems operated in
external device control (EDC) mode, in conjunction with Linear
Research LR400/LR700 four-probe ac resistance bridges, allowing
measurements down to $1.8$~K. The electrical contacts were placed
on the samples in standard 4-probe geometry, using Pt wires
attached to a sample surface with either Epotex H20e or Ablebond
88-1 silver epoxy, cured at 120$^{\circ}$C for 30 minutes. The
configuration was $I||[100]$; $H||[001]$ for samples with
easy-axis magnetization, and $I||[100]$; $H||[010]$ for samples
with easy-plane magnetization. Resistivity estimates for some
samples were obtained from the resistance measurements by cutting
the samples into rectangular slabs and estimating their
cross-sections and voltage contact distances, but since many of
these samples are quite thin the uncertainty in the absolute value
of $\rho(T)$ is in the range of 20\%. The resistive transition
temperatures were estimated from jumps or peaks in $d\rho/dT$
which often behaves in a manner very similar to $C_m(T)$ and
$d(\chi T)/dT$ near antiferromagnetic transitions
\cite{fis68a,esc81a,wien00a}.

Residual resistivity ratios are defined in this work as
$RRR=\rho$(300~K)/$\rho$(1.8~K), and in many cases the obtained
value is a lower limit since there is still a finite slope in
$\rho(T)$ at 1.8~K. When measured in zero applied field, $\rho(T)$
for most samples was found to drop 1\% to 6\% below 4~K due to the
onset of superconductivity in very small droplets of Sn flux that
remained on some of the sample surface. To estimate $RRR$
correctly, this effect must be eliminated by applying a field
above 0.3~kOe to drive Sn into its normal state. Given that the
change in $\rho(T)$ is so small at the $T_c$ of Sn, we can rule
out the possibility of normal-state elemental Sn shorting out any
significant percentage of the sample and giving rise to a
spuriously high $RRR$. This is supported by the fact that there is
no correlation between the percentile drops and the $RRR$ values
(see table~\ref{tbl1}). For example, even in the extreme case of a
20\% drop for $R=$~Y (see inset fig.~\ref{fYLu1}c) the sample
$RRR$ did not result among the highest values in the series. The
percentile drops due to superconducting Sn are visible in the
magnetoresistance graphs at 2~K for most samples, and in this case
the data was normalized to R(1~kOe) instead of R(0) to avoid this
influence in the analysis as well.

\section{Characterization of the compounds}

\subsection{YFe$_2$Ge$_2$ and LuFe$_2$Ge$_2$}

Prior to this work, the only reported properties of YFe$_2$Ge$_2$
were its crystal structure and cell parameters\cite{vent96a}.
LuFe$_2$Ge$_2$ is a previously unreported compound. Our
refinements of the x-ray diffraction pattern for YFe$_2$Ge$_2$
resulted in $a=3.968(2)$~\AA, $c=10.463(1)$~\AA,
$V=164.82(2)$~\AA$^3$. LuFe$_2$Ge$_2$ resulted in the same
$I4/mmm$ space group as most other {R}T$_2$X$_2$ compounds, with
$a=3.914(1)$~\AA, $c=10.395(1)$~\AA, $V=159.27(2)$~\AA$^3$.

\begin{figure}[thb]
\includegraphics[angle=0,width=33pc]{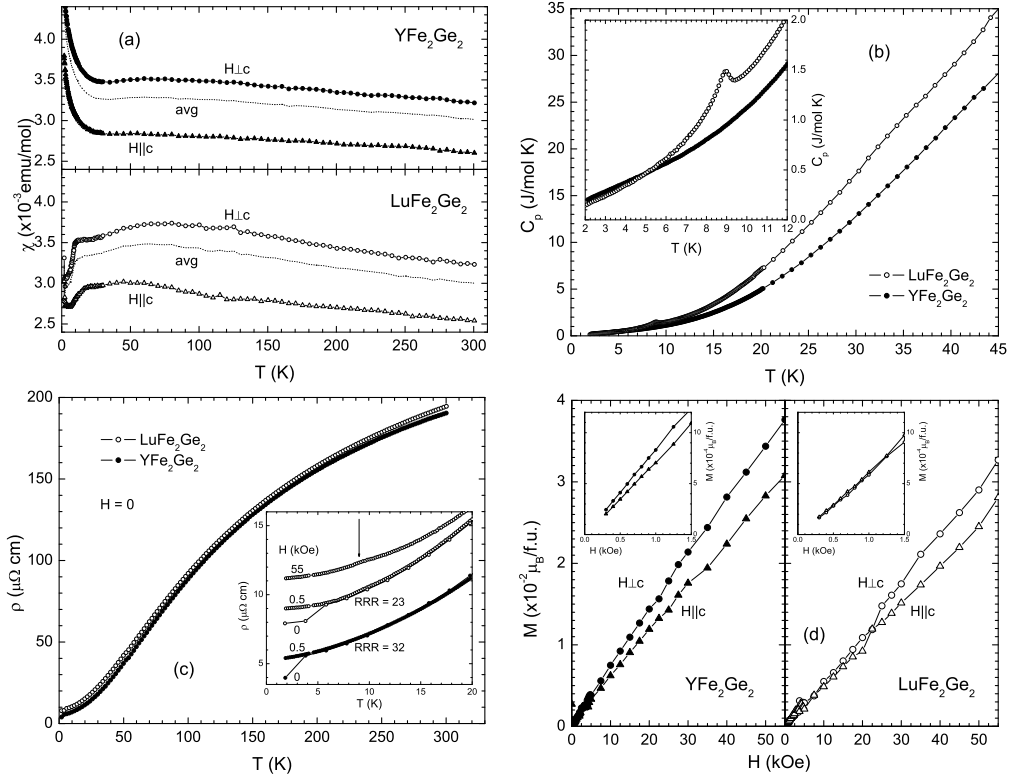}
\caption{\label{fYLu1} Measurements on YFe$_2$Ge$_2$ and
LuFe$_2$Ge$_2$ single crystals. (a) Anisotropic susceptibility at
$H=1$~kOe and polycrystalline averages. (b) Heat capacity at
$H=0$. The inset details the low temperature region. (c)
Resistivity at $H=0$. The inset shows low temperature resistivity
for $H=0$ and $H=0.5$~kOe of both compounds, plus $H=55$~kOe for
LuFe$_2$Ge$_2$. (d) Magnetization isotherms at $T=2$~K. The insets
detail the low-field region.}
\end{figure}

The magnetic behavior of both compounds is quite similar.
Figure~\ref{fYLu1}a shows the magnetic susceptibilities $\chi(T)$
at $H=1$~kOe which are, to a first approximation,
Pauli-paramagnetic and slightly anisotropic with
$\chi_{ab}(T)>\chi_c(T)$. This anisotropy is opposite to the case
of non-magnetic $R$Ni$_2$Ge$_2$ crystals\cite{bud99a}, and the
average molar susceptibility for the T=Fe crystals
($\chi_0\sim0.003$~emu/mol) is also about an order of magnitude
higher than for the T=Ni ones. This is a notably large $\chi_0$
and, as will be discussed below, must be taken into account when
analyzing the Curie tails of moment-bearing members of the series.
At low temperatures, the YFe$_2$Ge$_2$ samples showed a small
upturn probably due to a small amount of magnetic impurities (for
example, the tail would be equivalent to that produced by a
contamination of 0.1\% Gd). In contrast, the low temperature
behavior of LuFe$_2$Ge$_2$ showed a small but well marked anomaly
at 9~K (we will discuss this feature after presenting the other
measured properties in this compound where anomalies also appeared
at this temperature).

Figure~\ref{fYLu1}b shows the measured specific heats of both
compounds. $C_p(T)$ increases monotonically between 2~K and 45~K
for YFe$_2$Ge$_2$, whereas a peak centered at 9~K occurs for
LuFe$_2$Ge$_2$ (see inset), accompanying the reduction in
susceptibility. When plotted as $C_p/T~vs.~T^2$ a linear region is
observed between 11~K and 20~K, which extrapolates to
$\gamma\sim60$~mJ/mol~K$^2$ for both compounds. This is a rather
high value for the electronic specific heat coefficient, about 5
times larger than the values obtained for $R$Ni$_2$Ge$_2$, but
comparable to the value of 37~mJ/mol~K$^2$ estimated by Ebihara
\textit{et al.}\cite{ebih95a} in single crystal LaFe$_2$Ge$_2$,
and consistent with the larger $\chi_0$ terms found in the moment
bearing compounds (see table~\ref{tbl1}). Below 11~K,
$C_p^{(Lu)}/T$ is obviously non-linear due to the transition at
9~K, and $C_p^{(Y)}/T$ has a small upward deviation from
linearity, of unknown origin. The estimated Debye temperatures
from the slope of the same linear region mentioned above are
$\Theta_D=280$~K and $\Theta_D=240$~K for $R=$~Y and Lu
respectively.

The in-plane resistivity behaviors of YFe$_2$Ge$_2$ and
LuFe$_2$Ge$_2$ are almost identical, as shown in
figure~\ref{fYLu1}c. Starting at $\rho\sim200~\mu\Omega$~cm at
room temperature, they decrease monotonically with a broad
curvature upon cooling, followed by an inflexion point around
70~K, and finally tends to saturate at the lowest temperatures.
The inset shows the low temperature behavior for $H=0.5$~kOe,
where $\rho(T)$ reaches 5.4 and 9.0~$\mu\Omega$~cm at 1.8~K,
resulting in $RRR$ values of 32 and 23 for $R=$~Y and Lu
respectively, in good agreement with those obtained for $R=$~La
and Ce single crystals\cite{ebih95a} and annealed
ingots\cite{budk99a}. These values are almost 10 times larger when
compared to the T=Ni crystals, and can be understood as a combined
effect of a residual resistivity $\rho(0)$ 3-4 times smaller, with
resistivity at room temperature 2-3 times larger. The former
feature demonstrates the high crystallographic quality of the T=Fe
crystals, and the latter points to a characteristic effect of the
T=Fe compounds above $\sim70$~K, which increases electron
scattering and results in the broad curvature up to 300~K. The
inset of fig.~\ref{fYLu1}c also shows $\rho(T)$ at 55 kOe for
LuFe$_2$Ge$_2$, where a very subtle anomaly around 9~K is
observed.

The field-dependent magnetizations of YFe$_2$Ge$_2$ and
LuFe$_2$Ge$_2$ at 2~K are weakly anisotropic and essentially
linear below $H=55$~kOe (fig.~\ref{fYLu1}d). The insets show how
linearity is maintained down to 0.3~kOe and essentially
extrapolate to the origin (at 2~K the Sn droplets become
superconducting just below 0.3~kOe, leading to small diamagnetic
net response). This behavior demonstrates the absence of secondary
ferromagnetic phases in these samples.

The 9~K transition observed in LuFe$_2$Ge$_2$ appears to be a
robust feature which is very likely an intrinsic property of this
compound. It was observed as a loss of about 10\% in the
susceptibility of two different samples from different growth
batches, and was little affected by applied fields up to 55~kOe.
It is accompanied by a peak in heat capacity which amounts to
$\Delta C_p/C_p\sim20$\%, with transition width $\Delta
T_o/T_o\sim9$\%, and a subtle feature in the in-plane resistivity.
One possible explanation is the formation of a spin- or
charge-density wave which would abruptly decrease the
susceptibility by opening a gap in part of the Fermi surface.
Using the $\gamma$ value obtained above, we estimate $\Delta
C_p/\gamma T_o\sim0.46$, about 30\% of the BCS value of 1.43
expected in the mean field approximation\cite{kwok89a}, which
would imply a significant nesting effect on the Fermi surface. If
so, the nesting would most likely be along the [001] direction
given the small change in $\rho(T)$ around 9~K for $I\perp[001]$.
But obviously these are only rough estimates and further
investigation, including out-of-plane resistivity measurements,
will be required to confirm or refute this hypothesis.

\subsection{PrFe$_2$Ge$_2$}

Earlier
investigations\cite{mala93a,mali76a,leci83a,szyt83a,szyt90a,blai95a}
in PrFe$_2$Ge$_2$ ingots and powders have reported that this
compound is an antiferromagnet with $T_N\sim14$~K, and a second
transition temperature at $\sim9$~K which separates two different
ordering regimes: a low-temperature AFII-type magnetic structure
consisting of ferromagnetic layers of Pr moments aligned along the
$c$-axis in a $++--$ pattern, and a higher temperature
incommensurate magnetic structure where the moments remain aligned
with the $c$-axis but with an amplitude modulated sinusoidally
along the same axis. The ordered magnetic moment was found close
to the expected value of 3.2~$\mu_B$ for Pr$^{3+}$.
Field-dependent magnetization measurements in the ordered state
also showed a single metamagnetic transition near 15~kOe.

\begin{figure}[thb]
\includegraphics[angle=0,width=33pc]{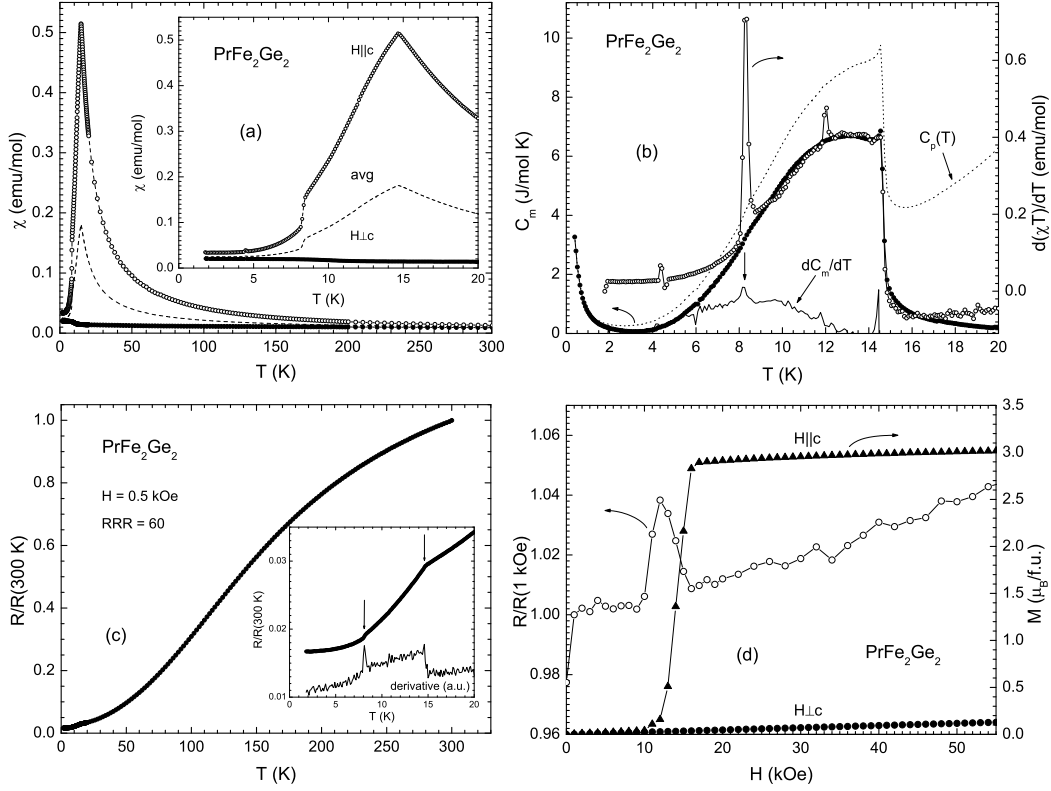}
\caption{\label{fPr1} Measurements on PrFe$_2$Ge$_2$ single
crystals. (a) Anisotropic susceptibility at $H=1$~kOe and
polycrystalline average. (b) Magnetic heat capacity at $H=0$
(solid symbols), its derivative (solid line) and $d(\chi T)/dT$
from the susceptibility average (open symbols). The dotted line
shows the raw heat capacity data. (c) Normalized resistance at
$H=0.5$~kOe. The inset details the low temperature region (left
scale) and $d\rho(T)/dT$ (arbitrary units). (d) Magnetization
isotherms at $T=2$~K (solid symbols) and normalized
magnetoresistance (open symbols).}
\end{figure}

The susceptibility at $H=1$~kOe of single crystal PrFe$_2$Ge$_2$
is strongly anisotropic (figure~\ref{fPr1}a). For $H||c$ the
susceptibility increases strongly as the sample is warmed,
presenting two transitions marked by a sharp jump at the lower one
and a well defined peak at the higher one, after which $\chi_c(T)$
decreases monotonically up to 300~K. For $H\perp{c}$ the
susceptibility is much smaller and decreases monotonically from
1.8~K, with a larger slope between the two aforementioned
transition temperatures. Analysis of the peaks in $d(\chi T)/dT$
(fig.~\ref{fPr1}b) places the two transition temperatures at 8.3~K
and 14.6~K respectively (as discussed in the experimental methods
section, the features near $T=4.2$ and 12~K are spurious, can be
seen in many of the $d(\chi T)/dT$ data sets, and will not be
discussed further). Fitting the polycrystalline average $\chi(T)$
above 20~K yields $\mu_{eff}=3.5~\mu_B$/f.u., close to the free
ion value of 3.57 $\mu_B$/f.u. for Pr$^{3+}$.

Specific heat measurements (fig.~\ref{fPr1}b) show a very sharp
rise as the sample is cooled below 15~K, which peaks at 14.6~K
marking the transition between disordered and ordered magnetic
states, and then a broad curvature with only a subtle change in
slope at 8.2~K (which nonetheless appears as a peak in the
derivative of this curve) that marks the transition between
different magnetically ordered phases. Below 3.2~K the specific
heat begins to rise strongly again, probably due to a nuclear
Schottky effect. Neglecting this rise, the estimated magnetic
entropy reaches $\sim4.3$~J/mol~K, somewhat short of
R$ln2=5.75$~J/mol~K, most likely due to an overestimate of the
lattice contribution when obtaining $C_m(T)$.

Normalized resistance measurements at $H=0.5$~kOe
(fig.~\ref{fPr1}c) show the same general behavior as those of the
non-magnetic compounds, but with $RRR=60$ and two well marked
changes in slope at 8.2~K and 14.6~K, as a consequence of changes
in the spin-disorder scattering regimes for different magnetic
phases.

Field-dependent magnetization at $T=2$~K shows an almost perfectly
linear behavior for $H\perp{c}$, reaching a value of
0.13~$\mu_B$/f.u. at 55~kOe. For $H||c$ the behavior is initially
similar, with a slope about 1.8 times larger, followed by a jump
of 2.9 $\mu_B$/f.u between 10 and 16~kOe consistent with a
spin-flop transition of the Pr moments, then adopts a similar
slope as for $H\perp{c}$, reaching a value of 3.0~$\mu_B$/f.u at
55~kOe. We found no evidence of a transition at 1 kOe claimed in a
previous work on polycrystals\cite{ivan92a}, which was more likely
an extrinsic effect due to small ferromagnetic impurities.
Normalized magneto-resistance at 2~K is small and positive,
showing a sharp peak during the spin-flop transition, and finally
resumes a featureless positive slope reaching about 8\% increase
at 90~kOe.

\subsection{NdFe$_2$Ge$_2$}

Earlier works on NdFe$_2$Ge$_2$ ingots and
powders\cite{feln75a,szyt83a} have shown that its magnetic
behavior is similar to PrFe$_2$Ge$_2$. It was claimed to order
antiferromagnetically with $T_N\sim13$~K, and to present the same
AFII-type magnetic structure with moments along the $c$-axis.

\begin{figure}[thb]
\includegraphics[angle=0,width=33pc]{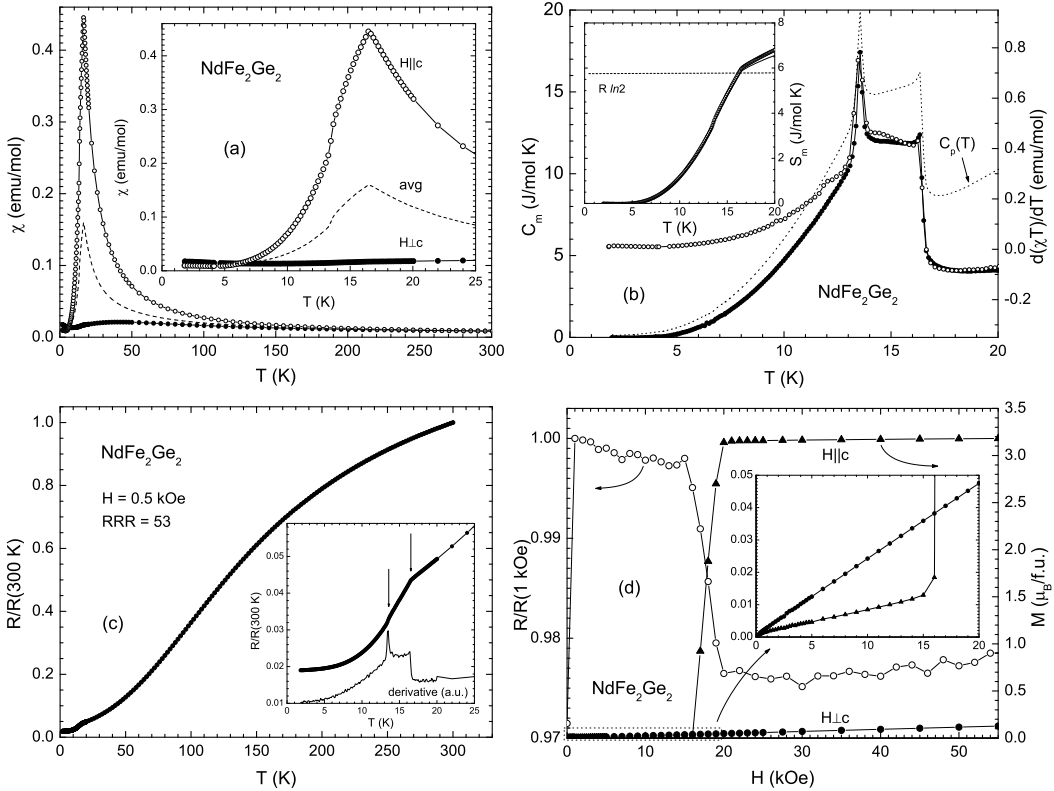}
\caption{\label{fNd1} Measurements on NdFe$_2$Ge$_2$ single
crystals. (a) Anisotropic susceptibility at $H=1$~kOe and
polycrystalline average. (b) Magnetic heat capacity at $H=0$
(solid symbols) and $d(\chi T)/dT$ from the susceptibility average
(open symbols). The dotted line shows the raw heat capacity data.
The inset shows the magnetic entropy as calculated from our
standard procedure (open circles), and the solid line is the
result obtained after direct subtraction of the LuFe$_2$Ge$_2$
specific heat. (c) Normalized resistance at $H=0.5$~kOe. The inset
details the low temperature region (left scale) and $d\rho(T)/dT$
(arbitrary units). (d) Magnetization isotherms at $T=2$~K (solid
symbols) and normalized magnetoresistance (open symbols). The
inset details the low field region.}
\end{figure}

The temperature dependent susceptibility of single crystal
NdFe$_2$Ge$_2$ is strongly anisotropic (fig.~\ref{fNd1}a) and
indeed similar to PrFe$_2$Ge$_2$ in general terms. $\chi_c(T)$
initially decreases slightly as temperature is raised from 1.8~K,
then rises fast and shows two transitions marked by a jump and a
sharp peak, after which it decreases in a roughly Curie-Weiss like
manner. $\chi_{ab}(T)$ is initially larger than $\chi_c(T)$ at the
lowest temperatures, decreasing below $\chi_c(T)$ as the
temperature is raised, and then exhibits a very broad peak
centered at around 40~K, indicating a CEF level splitting within
this thermal energy range. Analysis of the peaks in $d(\chi T)/dT$
(fig.~\ref{fNd1}b) places the two transition temperatures at
13.5~K and 16.4~K respectively, so the previously reported
transition observed in ingots may actually have been the lower
one. Fitting $\chi(T)$ above 20~K yields
$\mu_{eff}=3.4~\mu_B$/f.u., close to the free ion value of 3.61
$\mu_B$/f.u. for Nd$^{3+}$.

The magnetic specific heat of NdFe$_2$Ge$_2$ shows a sharp rise
which peaks at 16.4~K and, unlike PrFe$_2$Ge$_2$, the lower
transition is also marked by a sharp peak at 13.5~K
(fig.~\ref{fNd1}b). After this lower peak $C_m$ drops
monotonically down to the lowest measured temperature of 2~K, with
no indication of any further features. The total magnetic entropy
accumulated up to the 16.4~K transition is approximately
R\textit{ln}2, and continues to rise slowly above this
temperature, consistent with the presence of other CEF levels
contributing in this region (as discussed in the experimental
methods section, we used two different methods to account for the
non-magnetic component of the specific heat for these data. There
is virtually no difference for $T<T_N$ in this case).

The high temperature behavior of the normalized resistance at
$H=0.5$~kOe (fig.~\ref{fNd1}c) is similar to PrFe$_2$Ge$_2$, and
the low temperature region (inset) shows two well marked changes
in slope at 16.4~K and 13.5~K, followed by a lower temperature
levelling-off which results in $RRR=53$. The derivative $d\rho/dT$
makes the two transitions even clearer and shows a temperature
dependence that is very similar to that seen for $C_m(T)$ and
$d(\chi T)/dT$.

The field-dependent magnetization at 2~K is shown in
(fig.~\ref{fNd1}d). The planar magnetization is almost perfectly
linear with field and has a small slope, reaching
0.11~$\mu_B$/f.u. at 55~kOe. The axial magnetization begins rising
with a smaller slope than the planar one, but then undergoes a
jump of 3.1~$\mu_B$/f.u. between 15.5 and 20~kOe, then essentially
saturates, reaching 3.2~$\mu_B$/f.u. at 55~kOe. In a similar
manner, the normalized magneto-resistance starts with a small
negative slope as the field is increased, followed by a sharp 2\%
drop between 15 and 20~kOe accompanying the spin-flop transition
seen in magnetization, and finally a small positive slope which
persists up to 90~kOe consistent with saturation of the Nd
moments.

\subsection{SmFe$_2$Ge$_2$}

SmFe$_2$Ge$_2$ has been very little explored until now. Only the
crystal structure and lattice parameters were
determined\cite{bara90a}, $^{57}$Fe M\"{o}ssbauer spectroscopy was
performed at 77~K and 300~K\cite{bara90a}, and its melting
temperature was measured\cite{moro97a,moro98a}. To the best of our
knowledge, no characterization of its low-temperature properties
has been previously reported.

\begin{figure}[thb]
\includegraphics[angle=0,width=33pc]{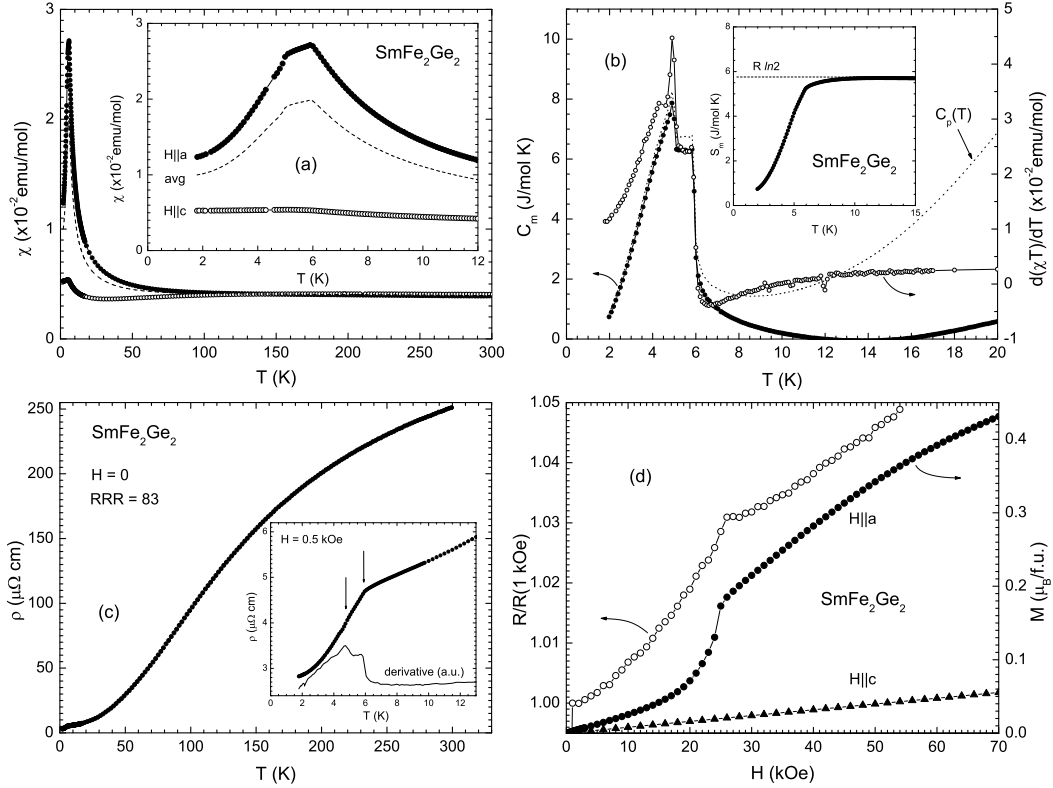}
\caption{\label{fSm1} Measurements on SmFe$_2$Ge$_2$ single
crystals. (a) Anisotropic susceptibility at $H=1$~kOe and
polycrystalline average. (b) Magnetic heat capacity at $H=0$
(solid symbols) and $d(\chi T)/dT$ from the susceptibility average
(open symbols). The dotted line shows the raw heat capacity data.
The inset shows the magnetic entropy. (c) Resistivity at
$H=0.5$~kOe. The inset details the low temperature region (left
scale) and $d\rho(T)/dT$ (arbitrary units). (d) Magnetization
isotherms at $T=2$~K (solid symbols) and normalized
magnetoresistance (open symbols).}
\end{figure}

Our susceptibility measurements (fig.~\ref{fSm1}a) showed
anisotropic behavior at temperatures up to about 120~K, with
$\chi_a(T)$ being significantly larger than $\chi_{c}(T)$ in this
region. $\chi_a(T)$ increases rapidly with increasing temperature,
and shows a well-marked double transition somewhat similar to
those observed in the Pr and Nd compounds, very likely indicating
antiferromagnetic ordering with moments aligned in or close to the
\textit{ab}-plane and two distinct magnetic structures. Above the
higher transition temperature $\chi_a(T)$ drops, but in a
distinctly non-Curie-Weiss manner. $\chi_c(T)$ initially rises
until the lower transition temperature is reached, then drops into
a broad minimum centered at 33~K, and finally rises again until
towards $\chi_a(T)$. The transition temperatures determined from
the peaks in $d(\chi T)/dT$ are 4.9~K and 5.9~K
(fig.~\ref{fSm1}b). Both curves assume essentially the same
levelled value of $\sim0.004$~emu/mol above $\sim120$~K, very
similar to that of the $R$=Y and Lu compounds, indicating that the
Sm ions essentially become non-magnetic either due to thermal
population of the upper Hund's rule multiplet or shift in the
valency from 3+ towards 2+ with increasing temperature. Similar
magnetic behavior was previously observed in SmNi$_2$Ge$_2$ single
crystals\cite{bud99a}. Fitting $\chi(T)$ above 20~K yields
$\mu_{eff}=0.46~\mu_B$/f.u., significantly smaller than the free
ion value of 0.84 $\mu_B$/f.u. for Sm$^{3+}$.

The magnetic specific heat (fig.~\ref{fSm1}b) rises sharply on
cooling through the upper transition temperature of 5.9~K, then
displays a plateau followed by a peak at 4.9~K and roughly linear
drop down to 2~K. The accumulated magnetic entropy at 5.9~K is
very close to R$ln2$ and then essentially levelled up to 20~K,
indicating a ground state doublet well separated in energy from
other excited levels.

The zero-field resistivity of SmFe$_2$Ge$_2$ (fig.~\ref{fSm1}c)
starts at $250~\mu\Omega$~cm and drops as the previously reported
members until a sharp change in slope marks the decrease of
spin-disorder scattering at 5.9~K. The lower transition at 4.9~K
isn't readily seen, but appears as a peak centered at 4.8~K in the
derivative plot. For this sample $RRR=83$ is one of the highest
values obtained in the series.

Field-dependant magnetization at $T=2$~K (fig.~\ref{fSm1}d) shows
a broad metamagnetic transition ending at 25~kOe for $H||a$, and
then continues to rise, reaching $0.43~\mu_B$/f.u. at the highest
measured field of 70~kOe. The finite slope at this field indicates
that the magnetization is not yet saturated. Magnetization for
$H||c$ is linear with field and reaches $0.055~\mu_B$/f.u. at
70~kOe. Magnetoresistance is positive and shows a feature around
25~kOe, marking the metamagnetic transition.

\subsection{GdFe$_2$Ge$_2$}

GdFe$_2$Ge$_2$ is one of the members that has been receiving
special attention in recent years both by itself\cite{duon02a} and
as part of the GdT$_2$Ge$_2$ group\cite{duon00a,muld93a}, taking
advantage of the fact that the spherically symmetric, half-filled
$4f$ shell of Gd$^{3+}$ reduces the influence of hidden
anisotropies in polycrystals, simplifying the application of
molecular and crystal field models and derivation of relevant
parameters. The most recent work by Duong \textit{et
al.}\cite{duon02a} presents a detailed study on single-phase
annealed ingots and powders. Besides confirming the
antiferromagnetic ordering at $\sim11$~K reported in earlier
works\cite{feln75a,mali76a,feln78a}, their main results are a
deviation from Curie-Weiss behavior in $\chi^{-1}(T)$, a saturated
magnetization in $M(H)$ at 4.2~K that falls somewhat shy of the
expected value of $7\mu_B$/f.u., and a magnetic specific heat
behavior $C_m(T)$ that peaks at 9.6~K on cooling, followed by a
broad anomaly around 3~K. All these features were explained based
on a generalized molecular-field model where the Gd moments
interact with an itinerant-electron band.

\begin{figure}[thb]
\includegraphics[angle=0,width=33pc]{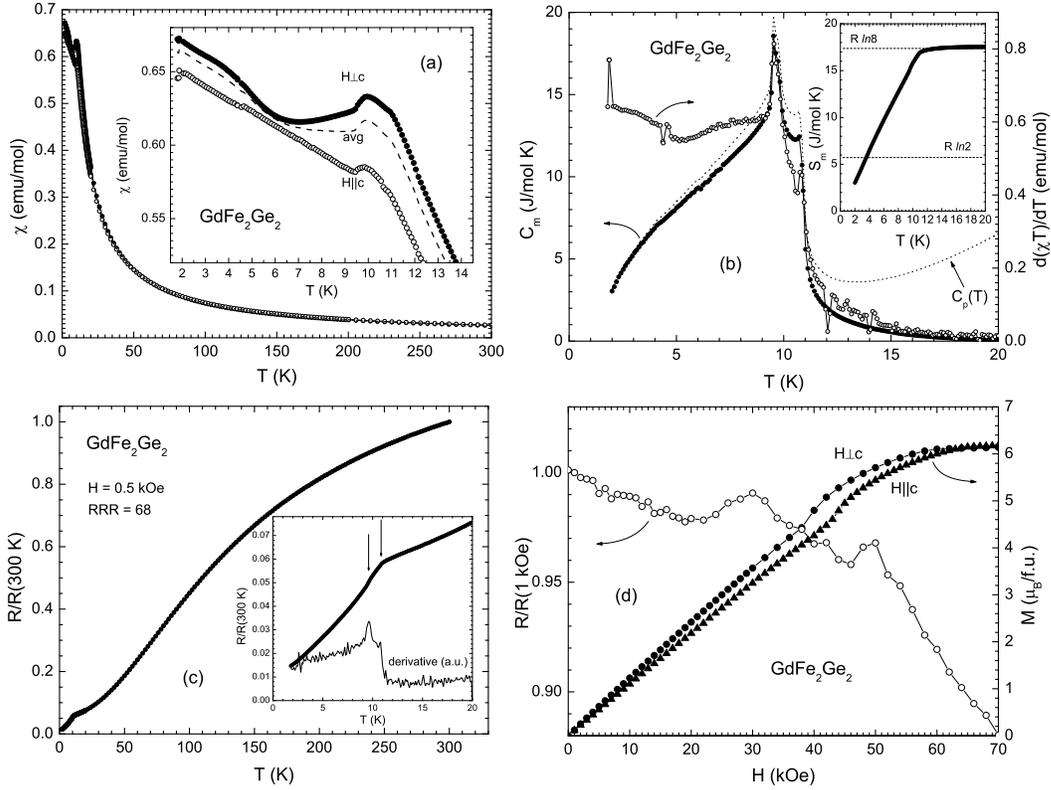}
\caption{\label{fGd1} Measurements on GdFe$_2$Ge$_2$ single
crystals. (a) Anisotropic susceptibility at $H=1$~kOe and
polycrystalline average. (b) Magnetic heat capacity at $H=0$
(solid symbols) and $d(\chi T)/dT$ from the susceptibility average
(open symbols). The dotted line shows the raw heat capacity data.
The inset shows the magnetic entropy. (c) Normalized resistance at
$H=0.5$~kOe. The inset details the low temperature region (left
scale) and $d\rho(T)/dT$ (arbitrary units). (d) Magnetization
isotherms at $T=2$~K (solid symbols) and normalized
magnetoresistance (open symbols).}
\end{figure}

As expected, the temperature dependent susceptibility at $H=1$~kOe
of single crystal GdFe$_2$Ge$_2$ in the paramagnetic state is
essentially isotropic (fig.~\ref{fGd1}a), with $\chi_{ab}(T)$
slightly higher than $\chi_c(T)$ (consistent with the non-magnetic
members of the series). The low temperature behavior is quite
rich, with the susceptibility undergoing several changes in slope
up to 11~K, after which it drops monotonically. By taking
$\chi_0=0.0026$~emu/mol into account there is no significant
deviation from Curie-Weiss behavior above 20~K, and the fit of
$\chi(T)$ yields $\mu_{eff}=7.6~\mu_B$/f.u., a little lower than
the expected value of $7.94~\mu_B$/f.u. for free-ion Gd$^{3+}$.
Analysis of $d(\chi T)/dT$ (fig.~\ref{fGd1}b) shows a large sharp
peak at 9.6~K and a smaller sharp peak/break at 10.8~K.

The magnetic specific heat of GdFe$_2$Ge$_2$ is also rich in
features (fig.~\ref{fGd1}b). It shows a sharp rise with a small
peak centered at 10.8~K on cooling, then continues rising towards
a larger peak at 9.6~K and finally displays a broad shoulder
around 3.0~K which has been observed in some other Gd compounds
such as GdBiPt\cite{can91a} and GdCu$_2$Si$_2$\cite{bouv91a} and
in the latter case it was attributed by the authors to a small
Zeeman splitting of the ground state octuplet in the ordered
state. The magnetic entropy reaches R\textit{ln}8 at the higher
transition temperature, as expected for a Gd ground state
octuplet. These features and values are all in excellent agreement
with ingots measured by Duong \textit{et al.}, except that for our
crystals we were able to resolve two very distinct transitions.

Normalized resistance as a function of temperature
(fig.~\ref{fGd1}c) clearly shows a loss of spin-disorder
scattering below 10.9~K, and a second, subtle change in slope at
9.6~K (easily seen as a peak in the derivative) marking the lower
transition. Again, there is a clear similarity between $d\rho/dT$
and $C_m(T)$ and $d(\chi T)/dT$. For this sample $RRR=68$.

The field-dependent magnetization at 2~K starts increasing
linearly, with the planar orientation slope slightly larger than
the axial one (fig.~\ref{fGd1}d). A clear change in behavior
occurs at 38~kOe and 43~kOe respectively, where magnetization
seems to show a small metamagnetic jump and then curves towards
saturation, reaching a value of $6.2~\mu_B$/f.u. at 70~kOe. This
value is essentially the same as that reported for powders at
70~kOe, but high field measurements on these show that the
magnetization continues to increase slowly until just below
$7~\mu_B$/f.u. at 400~kOe\cite{duon02a}. Normalized
magneto-resistance starts with a small negative slope, then shows
some rather ill-defined features between 30~kOe and 50~kOe, and
finally increases its slope reaching a 12\% drop at 70~kOe.

\subsection{TbFe$_2$Ge$_2$}

TbFe$_2$Ge$_2$ was first reported as having antiferromagnetic
ordering below 7.5~K and a metamagnetic transition at 15~kOe in
$M(H)$ measured at 4.2~K\cite{mali76a}. Neutron diffraction
experiments later described it as showing axial antiferromagnetic
ordering below 7.5~K with an incommensurate magnetic
structure\cite{pint85a}, although another neutron diffraction work
claimed that it was still paramagnetic at 4.2~K\cite{baze88a}, and
finally a more recent study that combined neutron diffraction with
magnetization measurements showed antiferromagnetic ordering with
$T_N=8.5$~K, $\mu(1.5$~K$)=7.68~\mu_B$/Tb, an incommensurately
modulated magnetic structure with two wave vectors, and a
metamagnetic transition at $H=11$~kOe for $T=4.2$~K\cite{szyt97a}.

\begin{figure}[thb]
\includegraphics[angle=0,width=33pc]{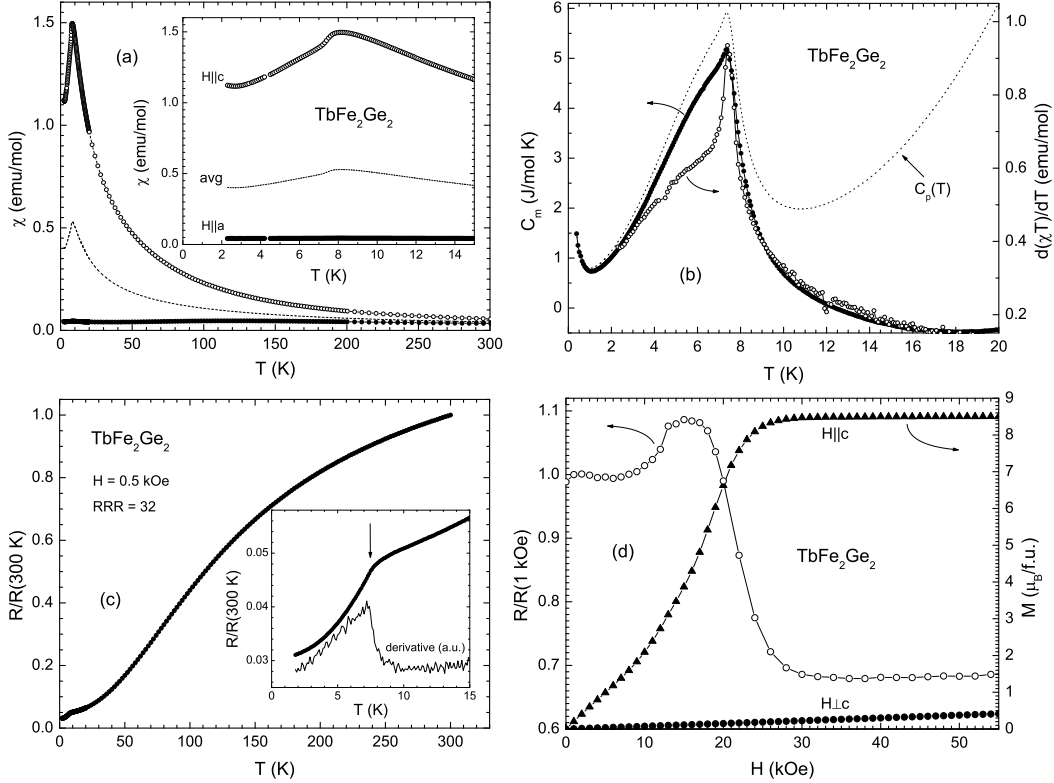}
\caption{\label{fTb1} Measurements on TbFe$_2$Ge$_2$ single
crystals. (a) Anisotropic susceptibility at $H=1$~kOe and
polycrystalline average. (b) Magnetic heat capacity at $H=0$
(solid symbols) and $d(\chi T)/dT$ from the susceptibility average
(open symbols). The dotted line shows the raw heat capacity data.
(c) Normalized resistance at $H=0.5$~kOe. The inset details the
low temperature region (left scale) and $d\rho(T)/dT$ (arbitrary
units). (d) Magnetization isotherms at $T=2$~K (solid symbols) and
normalized magnetoresistance (open symbols).}
\end{figure}

Strong anisotropies in the susceptibility exist for single crystal
TbFe$_2$Ge$_2$ (fig.~\ref{fTb1}a). The low temperature behavior of
$\chi_c(T)$ for $H=1$~kOe is dominated by a maximum at 8.2~K, with
a corresponding maximum in $d(\chi T)/dT$ at $T_N=7.4$~K
(fig.~\ref{fTb1}b). This may explain the 1~K discrepancy of
ordering temperatures reported in literature, since apparently the
authors reporting $T_N=8.5$~K used the maximum in $\chi(T)$ as
criterion. The susceptibility for $H||a$ is about 30 times
smaller, but presents the same peak at 8.2~K and upturn at the
lowest temperatures, which may indicate a small misalignment in
the orientation or that the ordered Tb moments form a small angle
with the \textit{c}-axis. It also shows a very broad maximum
centered at 120~K probably resulting from a CEF level split of
comparable thermal energy. Fitting $\chi(T)$ above 20~K yields
$\mu_{eff}=9.6~\mu_B$/f.u., close to the expected value of
$\mu_{eff}=9.72~\mu_B$/f.u. for Tb$^{3+}$.

The magnetic specific heat (fig.~\ref{fTb1}b) shows a more
broadened rise near the transition as those observed in other
members of the family. $C_m(T)$ reaches maximum value at 7.4~K,
then decreases with a broad curvature down to 1~K, after which it
starts rising again probably due to a nuclear Schottky effect.
This feature makes the estimation of the magnetic entropy
difficult. If the low temperature rise is ignored, the total
entropy up to $T_N$ reaches about 5~J/mol~K, close to R$ln2$ given
the qualitative nature of the low temperature extrapolation os our
$C_p(T)$ data. For this sample the procedure for estimation of
$C_m(T)$ did not work as well as with the others, since even with
the corrections $C_m(T)$ becomes spuriously negative above 13~K.

Normalized resistance at $H=0.5$~kOe (fig.~\ref{fTb1}c) has
essentially the same temperature dependence as the other members
of the family, and presents a single, rather broad change in
behavior near the ordering temperature due to loss of spin
disorder scattering. For this sample $RRR=32$, so the broadened
features discussed here are probably not resulting from poorer
sample quality.

The magnetization as a function of applied field at $T=2$~K is
strongly anisotropic (fig.~\ref{fTb1}d). For $H||c$ a broad upward
curvature exists between 6~kOe and 20~kOe, indicative of a
metamagnetic transition in this region. Above 30~kOe the
magnetization essentially saturates, reaching $8.5~\mu_B$/f.u. at
70~kOe, lower than the value of $9~\mu_B$/f.u. for Tb$^{3+}$
saturated moments. Application of the field parallel to the
\textit{a}-axis yields linear behavior, reaching $0.55~\mu_B$/f.u.
at 70~kOe. The magnetoresistance behavior starts practically flat,
then rises 9\% between 6 and 15~kOe, followed by a drop of almost
40\% between 15 and 30~kOe, after which it adopts a very small
positive slope up to 70~kOe.

All of the broadened features seen in different measurements of
the transition in TbFe$_2$Ge$_2$ indicate that for this compound
the behavior in this region could be a more washed-out version of
the double transitions seen in previous compounds, with the higher
magnetic phase existing in a very narrow temperature interval. On
the other hand TbFe$_2$Ge$_2$ may simply have a single magnetic
phase transition in low fields.

\subsection{DyFe$_2$Ge$_2$}

Few studies have been previously conducted on DyFe$_2$Ge$_2$. Its
crystal structure and lattice parameters were
determined\cite{ross78a,bara90a} and more recently Szytula
\textit{et al.} characterized its magnetic behavior\cite{szyt97a}.
The main reported results are: axial antiferromagnetic ordering
below 3.35~K, ordered magnetic moment of 7.68~$\mu_B$/f.u. at
1.5~K, and an incommensurately modulated magnetic structure
described by two wave vectors.

\begin{figure}[thb]
\includegraphics[angle=0,width=33pc]{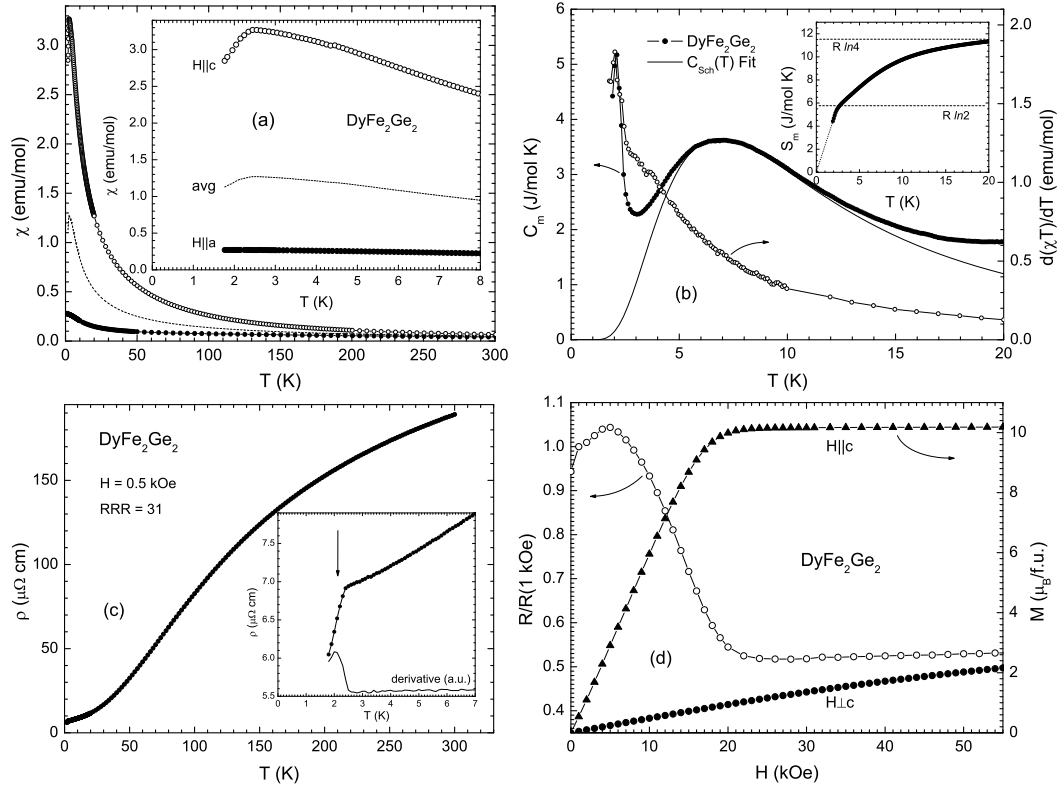}
\caption{\label{fDy1} Measurements on DyFe$_2$Ge$_2$ single
crystals. (a) Anisotropic susceptibility at $H=1$~kOe and
polycrystalline average. (b) Magnetic heat capacity at $H=0$
(solid symbols), $d(\chi T)/dT$ from the susceptibility average
(open symbols) and fit of eq.~\ref{eSch} (solid line). The inset
shows the magnetic entropy, where the dotted line results from
integration of the extrapolated $C_m(T)$ data to zero. (c)
Resistivity at $H=0.5$~kOe. The inset details the low temperature
region (left scale) and $d\rho(T)/dT$ (arbitrary units). (d)
Magnetization isotherms at $T=2$~K (solid symbols) and normalized
magnetoresistance (open symbols).}
\end{figure}

The magnetic susceptibility of single crystal DyFe$_2$Ge$_2$
(fig.~\ref{fDy1}a) is less anisotropic than TbFe$_2$Ge$_2$. The
axial susceptibility has a maximum at 2.5~K, after which it drops
rapidly. The planar susceptibility is similar, but with a
magnitude about 10 times smaller and ill-defined behavior below
2.5~K. Fitting $\chi(T)$ above 20~K yields
$\mu_{eff}=10.6~\mu_B$/f.u., essentially the expected value for
Dy$^{3+}$. The peak in $d(\chi T)/dT$ gives $T_N=2.1$~K
(fig.~\ref{fDy1}b) although there could be some uncertainty in
this value since we have only a few data points below 2.5~K. On
the other hand, specific heat and resistivity data (discussed
below) confirm this value.

\begin{figure}[htb]
\includegraphics[angle=0,width=33pc]{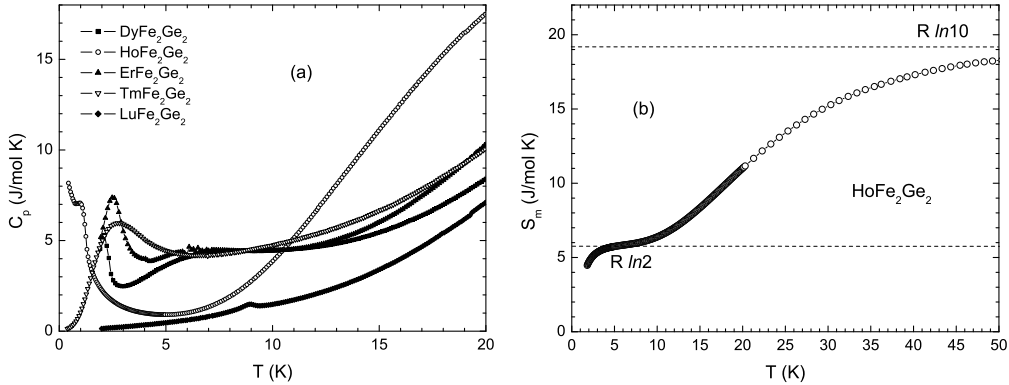}
\caption{\label{fExtra1} (a) As-measured heat capacities of
$R$Fe$_2$Ge$_2$ single crystals for $R=$~Dy-Tm and Lu. (b)
Estimation of the magnetic entropy of HoFe$_2$Ge$_2$, assuming
$S_m$(5~K)~=~R$ln2$.}
\end{figure}

For clarity, the as-measured specific heat $C_p(T)$ for
DyFe$_2$Ge$_2$ and also for $R=$~Ho-Tm will be shown separately in
figure\ref{fExtra1}a. The magnetic specific heat $C_m(T)$ of
DyFe$_2$Ge$_2$ (fig.~\ref{fDy1}b) is characterized by a broad
Schottky-like anomaly with local maximum at 6.8~K, below which a
sharp peak at 2.1~K marks the magnetic ordering and confirms the
value obtained from $d(\chi T)/dT$. The dashed curve in
fig.~\ref{fDy1}b is a fit to the Schottky contribution to magnetic
specific heat for a two-level system given by

\begin{equation}
C_{Sch}=R(\Delta/T)^2\frac{g_0}{g_1}\frac{e^{\Delta/T}}{[1+(g_0/g_1)e^{\Delta
/T}]^2}, \label{eSch}
\end{equation}

where $R=8314$~mJ/mol~K is the universal gas constant, $\Delta$ is
the energy gap in K, and $g_0/g_1$ gives the ratio between the
degeneracies of the lower and upper energy levels respectively.
From this fit we obtain $\Delta=16.5$~K and $g_0/g_1=1.0$. The
estimated magnetic entropy (inset) suffers from the lack of data
points below 1.9~K, but is close to R$ln2$ at $T_N$ and then
approaches R$ln4$ at 20~K. Given that Dy is a Kramers ion, these
data indicate that a pair of doublet ground states dominate the
low temperature properties of this compound.

Resistivity for $H=0.5~kOe$ (fig.~\ref{fDy1}c) starts at
$190~\mu\Omega$~cm at room temperature, and decreases like all
other members upon cooling until an abrupt change in slope occurs
around 2.1~K, marking a significant reduction of spin-disorder
scattering. At 1.8~K the resistivity is just above
$6~\mu\Omega$~cm and the resulting $RRR$ for this sample is 31,
although the sharp slope seen at this temperature indicates that
this value is to be considered as a lower limit.

The field-dependent magnetization of DyFe$_2$Ge$_2$ at 2~K for
$H||c$ results in a rapidly increasing magnetization up to 22~kOe,
above which it essentially saturates reaching $10.2~\mu_B$/f.u. at
55~kOe (fig.~\ref{fDy1}d). Applying the field in the basal plane
results in slightly sublinear behavior, reaching $2.2~\mu_B$/f.u.
at 55~kOe. Normalized magnetoresistance initially rises about 4\%
up to 5~kOe, then drops almost 50\% until the Dy moments saturate
at 22~kOe, and finally assumes a small positive slope with no
further features up to 90~kOe. These data are all consistent with
a metamagnetic transition that is complete for $H||c\sim22$~kOe.

\subsection{HoFe$_2$Ge$_2$}

An initial characterization of the magnetic properties of
HoFe$_2$Ge$_2$ was reported in 1997 by Szytula \textit{et al.} in
the same work that characterized the compounds with
$R$=Tb,Dy\cite{szyt97a}. They found no sign of ordering down to
1.5~K, but neutron diffraction measurements indicated signs of
long-range magnetic interactions that could lead to
antiferromagnetic ordering at some lower temperature. In contrast,
another work by Schobinger-Papamantellos \textit{et al.} reported
that their HoFe$_2$Ge$_2$ samples showed axial arrangement of the
Ho moments below 17~K, with wave vector (0.5,0.5,0). Short range
order effects were reported below 6~K, and the Ho moment at 1.5~K
was estimated as 6.6~$\mu_B$. Szytula \textit{et al.} have
recently performed new neutron diffraction experiments and found
the same collinear antiferromagnetic structure which however
disappears above 7~K, a short-range magnetic order which
disappears above 5~K, and Ho moment of 4.5~$\mu_B$. The conclusion
of the authors who performed these three studies was that
HoFe$_2$Ge$_2$ represents a rare case of coexisting long-range and
short-range orders, resulting from a competition between RKKY
interactions and CEF effects, and leading to a frustration of the
magnetic order.

\begin{figure}[thb]
\includegraphics[angle=0,width=33pc]{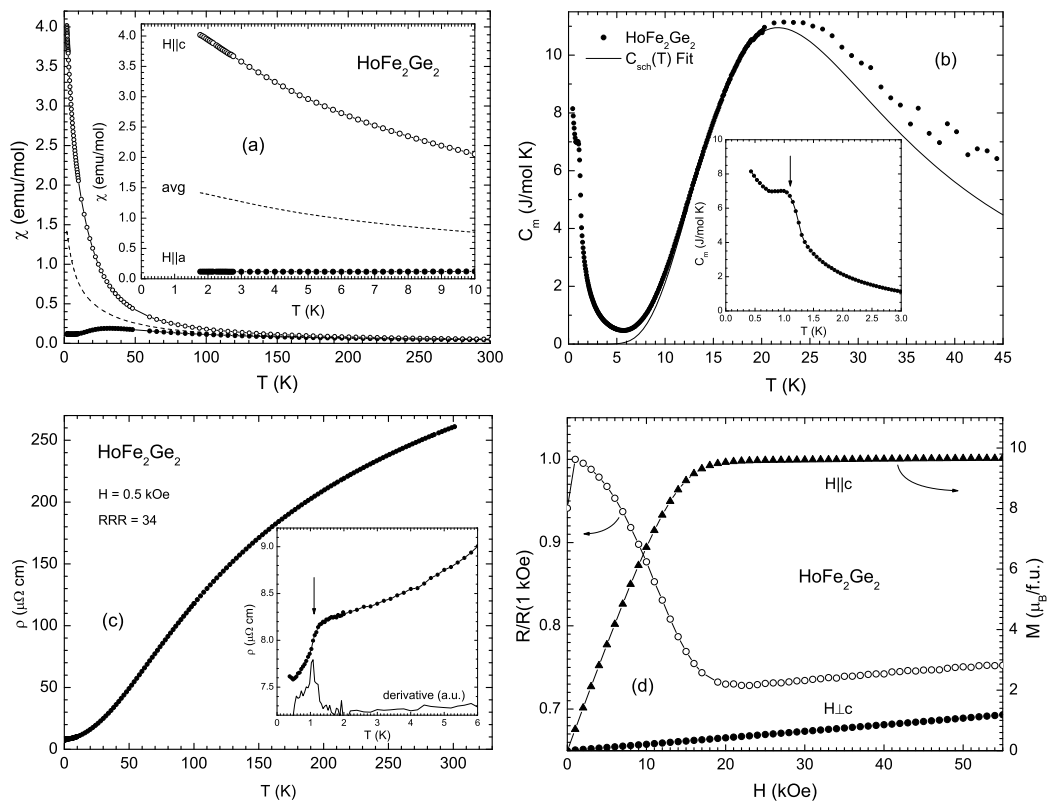}
\caption{\label{fHo1} Measurements on HoFe$_2$Ge$_2$ single
crystals. (a) Anisotropic susceptibility at $H=1$~kOe and
polycrystalline average. (b) Magnetic heat capacity at $H=0$
(solid symbols) and fit of eq.~\ref{eSch} (solid line). The inset
shows the low temperature region. (c) Resistivity at $H=0.5$~kOe.
The inset details the low temperature region (left scale) and
$d\rho(T)/dT$ (arbitrary units). (d) Magnetization isotherms at
$T=2$~K (solid symbols) and normalized magnetoresistance (open
symbols).}
\end{figure}

No transitions are observed in our susceptibility measurements on
HoFe$_2$Ge$_2$ single crystals above 1.8 K (fig.~\ref{fHo1}a). The
response is very anisotropic, with the axial susceptibility
featureless and much greater than in the planar one. $\chi_c(T)$
decreases monotonically, whereas $\chi_a(T)$ shows a local minimum
at about 6~K, followed by a broad maximum at about 30~K. Fitting
$\chi(T)$ above 20~K yields $\mu_{eff}=11.0~\mu_B$/f.u., slightly
above the expected value of $\mu_{eff}=10.6~\mu_B$/f.u. for
Ho$^{3+}$.

Specific heat measurements of HoFe$_2$Ge$_2$ reveal some
interesting features (fig.~\ref{fHo1}b), and the behavior is in
excellent quantitative agreement with that reported by
Schobinger-Papamantellos \textit{et al.} on annealed ingots. A
very large Schottky anomaly dominates the low temperature
behavior. The dashed line in fig.~\ref{fHo1}a represents
eq.~\ref{eSch} with $\Delta=66$~K and $g_0/g_1=0.23$. Below 5.6~K
$C_m(T)$ begins to rise very sharply again, and a small local
maximum close to 1~K can be clearly seen. These features indicate
that the heat capacity behavior below 6~K may be a convolution of
Ho nuclear Schottky effect with magnetic fluctuations of the
electronic moments, including antiferromagnetic ordering of the
latter. If this is the case, then the CEF level scheme should
consist of a ground state doublet and an 8-fold near-degenerate
manifold of CEF levels in the vicinity of $\Delta_{CEF}\sim60$~K.
The other possible scenario would be that the small peak is not
related to magnetic ordering of the Ho moments, and the ground
state is a non-magnetic singlet with a 4-fold near-degenerate
excited level manifold. The large nuclear Schottky feature
prevents a reliable quantitative analysis of the magnetic entropy
which would be very useful in distinguishing between these two
options, however, we will show below that the first scenario is
the more likely one. A simple test of consistency for this
hypothesis can be made by ignoring the data below the plateau
around 5~K and postulating that the accumulated magnetic entropy
up to this temperature has reached R$ln$2 due to the
antiferromagnetic transition. When this is done
(fig.~\ref{fExtra1}b), the magnetic entropy is seen to approach
R$ln$10 as the temperature approaches 60~K.

Further corroboration of a magnetically ordered ground state was
provided by temperature-dependent resistivity measurements at
0.5~kOe (fig.~\ref{fHo1}c), where we can observe a clear loss of
spin-disorder scattering below 1.1~K, which we believe is probably
associated with an antiferromagnetic transition temperature. It is
important to note that the 1.1~K feature is not related in any way
to Sn flux since, by measuring under zero applied field, the small
drop due to Sn superconductivity appears at $\sim$4~K while the
1.1~K drop remains intact. The high temperature behavior of
$\rho(T)$ is similar to all other members, reaching
$260~\mu\Omega$~cm at 300~K, and for this sample $RRR=34$.

Field dependent magnetization at 2~K (therefore above the
transition) rises rapidly up to 20~kOe for $H||c$ and remains
almost levelled thereafter, reaching $9.7~\mu_B$/f.u. at 55~kOe
(fig.~\ref{fHo1}d). Magnetization in the basal plane is small and
perfectly linear with field, reaching $1.19~\mu_B$/f.u. at 55~kOe.
Magnetoresistance measurements at 2~K show an almost 30\% drop
which accompanies the field-driven alignment of Ho magnetic
moments up to 20~kOe, then assumes a small positive slope up to
90~kOe in the high-field, saturated moments region.

The fact that we observe clear local moment behavior in $M(H)$ at
$T=2$~K indicates that the sample is not in a non-magnetic singlet
state with the first excited CEF states separated by $\sim60$~K.
The large magnetoresistance at $T=2$~K also supports this
observation. These data support the idea that the 1.1~K transition
is magnetic in character, arising from a ground state doublet, and
that there is a multiplet of 8 CEF levels at $\sim60$~K.

\subsection{ErFe$_2$Ge$_2$}

Except for the determination of its crystal structure and lattice
parameters\cite{bara90a}, melting temperature\cite{moro97a} and
for an early report stating that it is paramagnetic from room
temperature down to 4.2~K\cite{mali76a}, the properties of
ErFe$_2$Ge$_2$ have so far remained unexplored.

\begin{figure}[thb]
\includegraphics[angle=0,width=33pc]{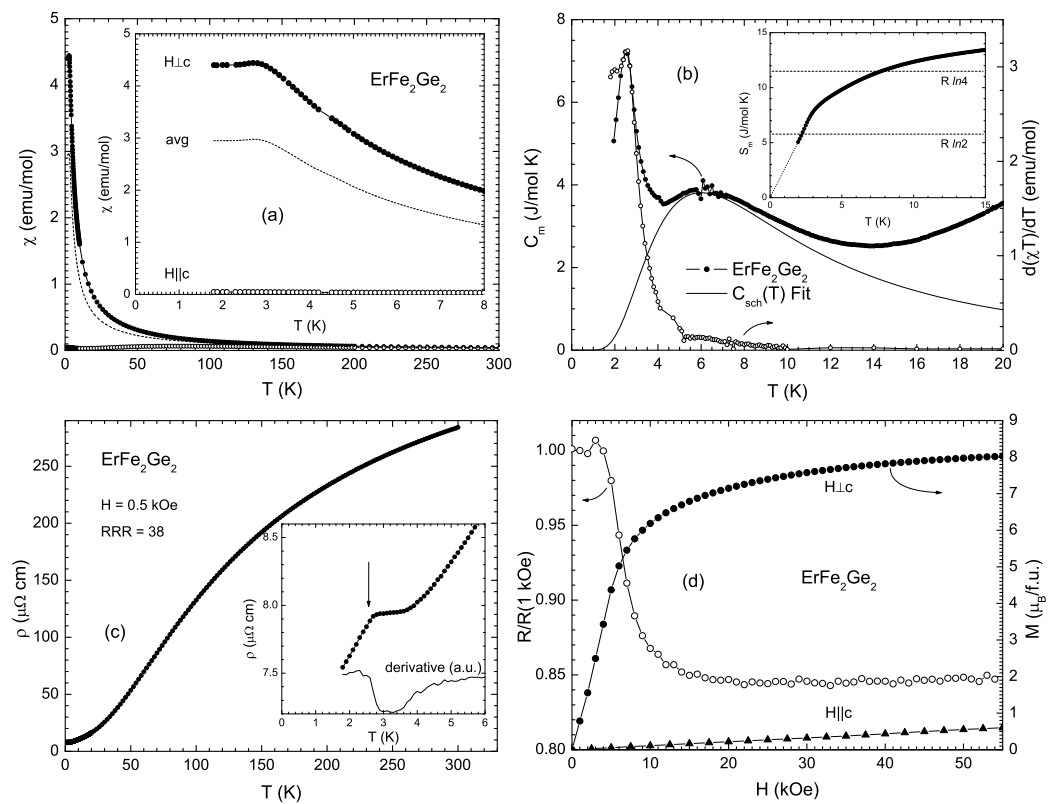}
\caption{\label{fEr1} Measurements on ErFe$_2$Ge$_2$ single
crystals. (a) Anisotropic susceptibility at $H=1$~kOe and
polycrystalline average. (b) Magnetic heat capacity at $H=0$
(solid symbols), $d(\chi T)/dT$ from the susceptibility average
(open symbols) and fit of eq.~\ref{eSch} (solid line). The inset
shows the magnetic entropy, where the dotted line results from
integration of the extrapolated $C_m(T)$ data to zero. (c)
Resistivity at $H=0.5$~kOe. The inset details the low temperature
region (left scale) and $d\rho(T)/dT$ (arbitrary units). (d)
Magnetization isotherms at $T=2$~K (solid symbols) and normalized
magnetoresistance (open symbols).}
\end{figure}

The magnetic behavior of single crystal ErFe$_2$Ge$_2$ is very
anisotropic (fig.~\ref{fEr1}a), with the moments aligning
perpendicular to the \textit{c}-axis (easy plane), consistent with
a change in sign of the Stevens coefficient $B_2^0$ in the CEF
Hamiltonian (discussed in the following section). $\chi_{ab}(T)$
is large and initially constant between 1.8 and 2.3~K, then rises
to a maximum at 2.7~K and drops monotonically up to room
temperature. $\chi_{c}(T)$ is very small, showing a minimum at
about 10~K followed by broad maximum at 75~K. Fitting $\chi(T)$
above 20~K yields $\mu_{eff}=9.5~\mu_B$/f.u., very close to the
free ion value of 9.58~$\mu_B$/f.u. for Er$^{3+}$. From the peak
in $d(\chi T)/dT$ we estimate $T_N=2.5$~K (fig.~\ref{fEr1}b).

The magnetic specific heat of ErFe$_2$Ge$_2$ shows a peak at 2.5~K
marking the antiferromagnetic transition, followed by a broad
Schottky-like maximum near 6.1~K, above which there still is a
significant amount of magnetic entropy probably due to the
presence of other CEF levels contributing in this region. The
dashed curve in fig.~\ref{fEr1}b is a fit of eq.\ref{eSch},
resulting in $\Delta=14.5$~K and $g_0/g_1=0.95$. The scenario
seems to be similar to that of DyFe$_2$Ge$_2$, except for the
inverted anisotropy and the rather high values of $C_m$ above the
Schottky peak. The magnetic entropy appears to be close to R$ln2$
near $T_N$ (inset fig~\ref{fEr1}b), but the overall behavior
doesn't allow a very reliable interpretation probably due to a
more complex upper CEF level scheme.

Resistivity for $H=0.5$~kOe is about $280~\mu\Omega$~cm at room
temperature (fig.~\ref{fEr1}c), decreases like all other members
of the series until levelling at about 4~K, then drops again
around 2.6~K marking the magnetically ordered state. At 1.8~K the
resistivity has been reduced to $7.5~\mu\Omega$~cm, corresponding
to $RRR=38$ which should be taken as a lower limit since the slope
is still high at this temperature.

Field dependent magnetization at 2~K and along the \textit{ab}
plane rises very rapidly and initially shows a small upward
curvature (fig.~\ref{fEr1}d), which indicates a metamagnetic
transition around 3~kOe. By 10~kOe the magnetization is at
$6.2~\mu_B$/f.u. and from there it continues to grow slowly,
reaching $8.1~\mu_B$/f.u. at 55~kOe but still with a positive
slope. This may indicate a canted ordering of the moments within
the basal plane or in-plane anisotropy, and further investigation
of the in-plane magnetic behavior will be required to clarify
these issues. The axial magnetization is very small and
essentially linear, reaching $0.6~\mu_B$/f.u. at 55~kOe.
Normalized magnetoresistance shows a small peak near 3~kOe, giving
support to the existence of a metamagnetic transisition, then
drops about 15\% and remains practically levelled up to 55~kOe.

\subsection{TmFe$_2$Ge$_2$}

\begin{figure}[thb]
\includegraphics[angle=0,width=33pc]{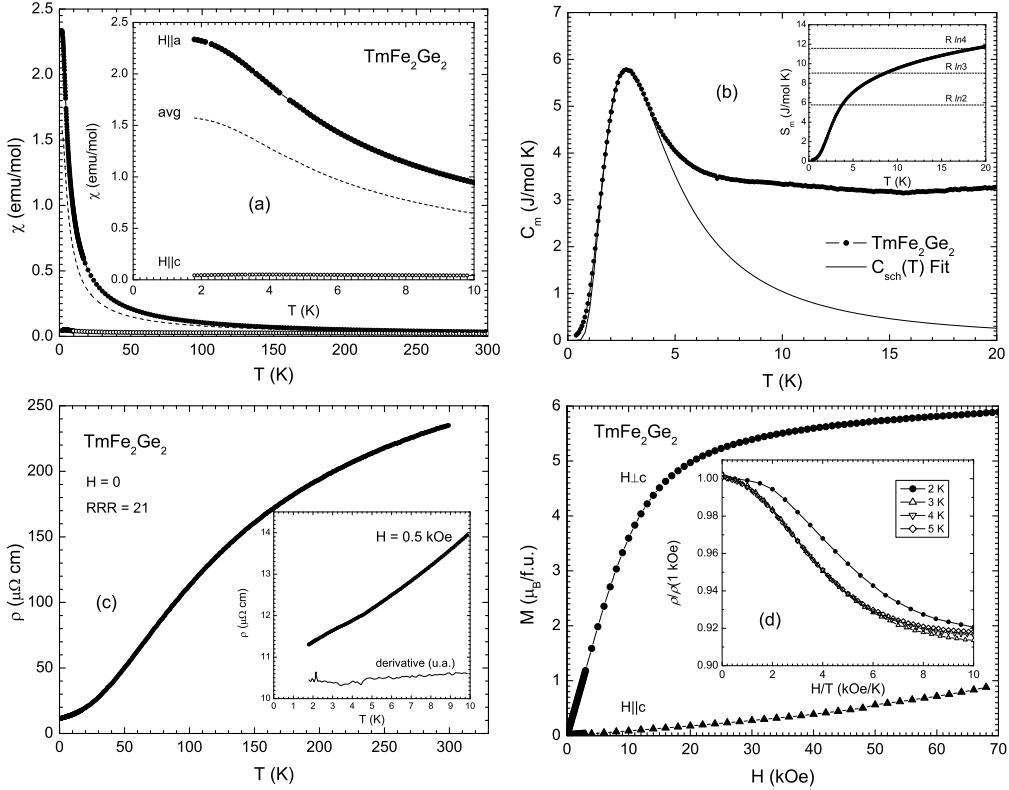}
\caption{\label{fTm1} Measurements on TmFe$_2$Ge$_2$ single
crystals. (a) Anisotropic susceptibility at $H=1$~kOe and
polycrystalline average. (b) Magnetic heat capacity at $H=0$
(solid symbols) and fit of eq.~\ref{eSch} (solid line). The inset
shows the magnetic entropy. (c) Resistivity at $H=0$. The inset
details the low temperature region (left scale) and $d\rho(T)/dT$
(arbitrary units). (d) Magnetization isotherms at $T=2$~K. The
inset shows normalized magnetoresistance $vs.~H/T$ for different
temperatures.}
\end{figure}

The magnetic behavior of single crystal TmFe$_2$Ge$_2$
(fig.~\ref{fTm1}a) is more anisotropic than ErFe$_2$Ge$_2$, with
the moments also aligning perpendicular to the \textit{c}-axis
(easy plane). $\chi_{a}(T)$ is large and decreases monotonically
above 1.8~K, although at the lowest temperatures there is a clear
change in curvature. $\chi_{c}(T)$ is very small and increases to
a maximum at 3.8~K, then drops monotonically. It is tempting to
attribute this to antiferromagnetic ordering, but the fact that it
appeared only in the hard-axis measurement warrants further
investigation. As will be shown below, this feature is most likely
associated with a crossover to a low temperature non-magnetic
singlet ground state. Fitting $\chi(T)$ above 20~K yields
$\mu_{eff}=7.8~\mu_B$/f.u., slightly larger than the free ion
value of 7.56~$\mu_B$/f.u. for Tm$^{3+}$.

The specific heat of TmFe$_2$Ge$_2$ was measured in the $^3$He
system to help clarify the sample behavior down to 0.39~K.
$C_m(T)~vs.~T$ shows a large but broadened peak centered at 2.1~K
(fig.~\ref{fTm1}b), which drops towards zero as the lowest
temperatures are reached. The shape is more representative of a
Schottky anomaly than a lambda-like ordering peak. Indeed, the
lower part fits well to eq.\ref{eSch} with $\Delta=7.1$~K and
$g_0/g_1=0.56$, and the accumulated entropy in this temperature
region is close to R$ln$3 (inset fig.~\ref{fTm1}b). This would be
consistent with a non-magnetic CEF singlet ground state and a
first excited state formed by a doublet or two close singlets,
allowing the appearance of magnetism by thermal population of
these first excited states, but above 7~K the level scheme should
have a more complicated distribution of levels (probably mostly
singlets with different energy separations) which would require a
more elaborate data modelling to fully describe the low
temperature behavior. Neglecting these higher states, the obtained
level scheme would help explain the curvature observed in
$\chi_{ab}(T)$ at the lowest temperatures, since the singlet
ground state and energy gap would lead to a levelled van
Vleck-like susceptibility for $T\ll\Delta$. In its simplest
modelling\cite{kittel4}, the product $\chi_{vv}\Delta$, where
$\chi_{vv}$ is the temperature-independent molar susceptibility,
depends only on the non-diagonal matrix element
$<$0$\mid$$\mu_z$$\mid$s$>$ connecting the singlet ground state
$0$ with the excited state $s$. If we assume that $\chi_{ab}(T)$
for TmFe$_2$Ge$_2$ will level at $\chi_{vv}\sim2.4$~emu/mol
(fig.\ref{fTm1}a), we obtain $\chi_{vv}\Delta\sim17$~K~emu/mol,
comparable to $\sim14$~K~emu/mol that can be estimated for some
well known thulium-based van Vleck paramagnets such as
TmSb\cite{rett75a} and TmH$_2$\cite{weiz96a}. It should be noted
that the expression above is derived for the particular case where
the CEF has removed all orbital degeneracies. It is also worth
mentioning that in some compounds, strong enough exchange
interactions can lead to magnetic ordering even when the CEF
ground state is a singlet. For example, TmCu$_2$Si$_2$ has been
reported\cite{stew88a,koza97a} as having two singlet states
separated by $\Delta=6.1$~K, and yet orders antiferromagnetically
at $T_N\sim2.8$~K. However, such an ordering should be accompanied
by a lambda-like peak in the magnetic specific heat as indeed
occurs with TmCu$_2$Si$_2$\cite{stew88a}, and our measurements
show no sign of any such feature down to 0.4~K in TmFe$_2$Ge$_2$.

Zero-field resistivity is about $235~\mu\Omega$~cm at room
temperature (fig.~\ref{fTm1}c), and behaves much like
LuFe$_2$Ge$_2$ upon cooling, showing no clear sign of loss of
spin-disorder scattering down to 1.8~K, at which point it is
slightly above $11~\mu\Omega$~cm, resulting in $RRR=21$.

Field dependent magnetization at 2~K and along the \textit{ab}
plane rises in a reversible Brillouin-like manner
(fig.~\ref{fTm1}d), reaching only $5.9~\mu_B$/f.u. at 70~kOe,
lower than the saturated moment of $7~\mu_B$/f.u. for Tm$^{3+}$.
The axial magnetization is small and seems essentially linear,
although high-field measurements in this orientation were
difficult due to the strong torque applied on the sample, arising
from $H\perp{M}$. Magnetoresistance was measured at $T=2$, 3, 4
and 5~K. For these temperatures the resistivity is seen to
decrease with $H$ rather rapidly until some field value where it
reaches a minimum, then assumes a small positive slope. As
expected for spin-disorder scattering by paramagnetic
moments\cite{fish97a} for $T\gg T_N$, the normalized resistivity
is seen to scale very well with $H/T$ for the measurements between
3~K and 5~K, but not for the measurements at 2~K, indicating that
at least some change in the scattering regime of the sample has
occurred at 2~K, as a consequence of Tm ions dropping into their
non-magnetic state.

\section{Trends in the series}

\begin{figure}[thb]
\includegraphics[angle=0,width=33pc]{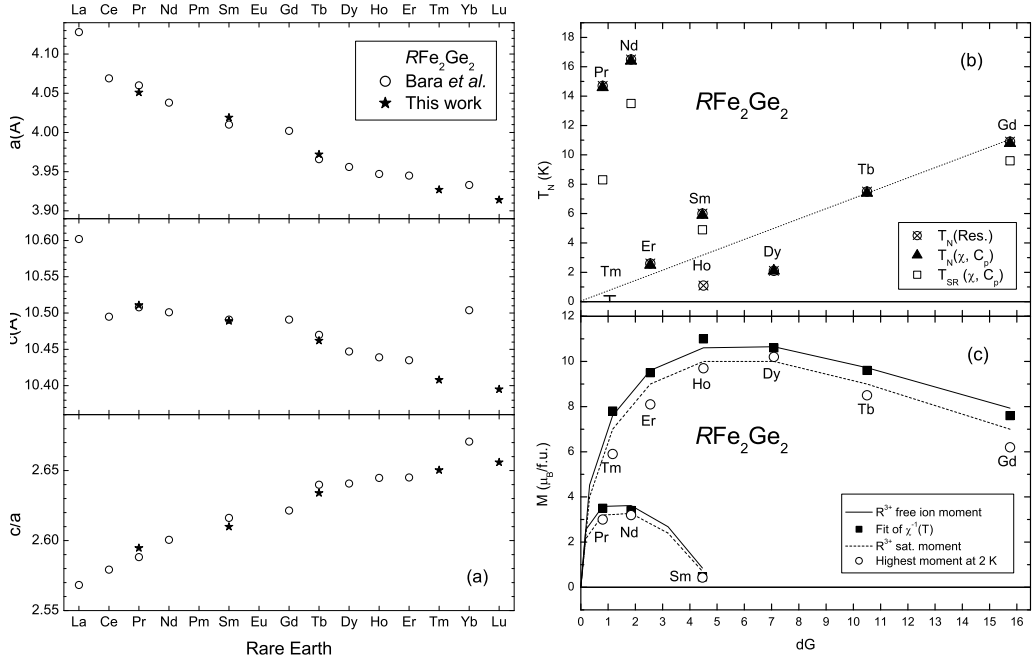}
\caption{\label{fTrends} (a) Lattice parameters obtained from
refinements of x-ray powder diffractograms on $R$Fe$_2$Ge$_2$
annealed ingots (Bara \textit{et al.}\cite{bara90a}) and some
selected single crystals (this work). (b) Scaling of the
$R$Fe$_2$Ge$_2$ transition temperatures with de Gennes factor
$(g_J-1)^2J(J+1)$. (c) Comparison of the expected and measured
values for the free ion moments and saturated moments.}
\end{figure}

When crystals are grown in a foreign-element flux it is always
good to check whether there is any significant substitution of the
phase elements by the flux element. In our case the most likely
scenario would be Sn substituting Ge within the structure. If this
substitution occurred at a relevant degree, we would expect to see
changes in the lattice parameters since Sn is larger than Ge. The
refined lattice parameters we obtained from powder x-ray
diffractograms on some crushed crystals(figure~\ref{fTrends}a) are
in excellent agreement with published values on arc-melted
ingots\cite{bara90a}, and further corroboration of phase purity is
given by the very low residual resistivities. The general trend of
the lattice parameters in this series is consistent with the
reduction in size of the rare earth 3+ ion with increasing atomic
number, leading to a small decrease of the unit cell volume but
also to a small elongation of the tetragonal unit cell. A notable
exception is YbFe$_2$Ge$_2$ where Yb is reported to be in a mixed
or intermediate valency state\cite{gros86a}, and this may lead to
changes in the phase diagram which resulted in our failure to grow
this compound out of Sn flux. Our other failed attempts were for
$R=$~La and Ce due to interference of a secondary cubic phase
(probably $R$Sn$_3$), and EuFe$_2$Ge$_2$ which apparently is not a
stable phase since there are no reports of successful synthesis of
this compound, not even in Felner and Nowik's systematic study of
EuT$_2$Ge$_2$ arc-melted ingots\cite{feln78a}.

\begin{table*}
\begin{tabular*}{32pc}[]{l|cccccccc}
\hline \hline
$Compound$ & $\alpha_J$ & Easy & $T_N$ & $T_{SR}$ & $\chi_0$ & $\mu_{eff}$ & $\mu_{HF}$ & $RRR$ \\
 & $(\times10^2)$ & orient. & (K) & (K) & (emu/mol) & ($\mu_B$) & ($\mu_B$) & \\
\hline
YFe$_2$Ge$_2$ & - & - & - & - & .0030 & - & .038 & 32 \\
PrFe$_2$Ge$_2$ & -1.05 & axial & 14.6 & 8.3 & .0053 & 3.5 & 3.0 & 60 \\
NdFe$_2$Ge$_2$ & -.643 & axial & 16.4 & 13.5 & .0042 & 3.4 & 3.2 & 53 \\
SmFe$_2$Ge$_2$ & +4.13 & planar & 5.9 & 4.9 & .0039 & .46 & .43 & 83 \\
GdFe$_2$Ge$_2$ & - & - & 10.8 & 9.6 & .0026 & 7.6 & 6.2 & 68 \\
TbFe$_2$Ge$_2$ & -1.01 & axial & 7.4 & - & .0058 & 9.6 & 8.5 & 32 \\
DyFe$_2$Ge$_2$ & -.635 & axial & 2.1 & - & .0043 & 10.6 & 10.2 & 31 \\
HoFe$_2$Ge$_2$ & -.222 & axial & 1.1 & - & .0024 & 11.0 & 9.7 & 34 \\
ErFe$_2$Ge$_2$ & +.254 & planar & 2.5 & - & .0025 & 9.5 & 8.1 & 38 \\
TmFe$_2$Ge$_2$ & +1.01 & planar & $<.4$ & - & .0049 & 7.8 & 5.9 & 21 \\
LuFe$_2$Ge$_2$ &- & - & - & - & .0030 & - & .033 & 23 \\
\hline \hline
\end{tabular*}

\caption{\label{tbl1} Summary of the measured properties of the
$R$Fe$_2$Ge$_2$ single crystals (except $\alpha_J$ obtained from
ref.~\cite{hut64a}). $T_N=$ Ne\'el temperature, $T_{SR}=$ second
magnetic transition, $\chi_0=$ temperature-independent
susceptibility term, $\mu_{eff}=$ effective moment, $\mu_{HF}=$
highest measured moment at 2~K, $RRR=R(300K)/R(1.8K)$. }
\end{table*}

Table~\ref{tbl1} summarizes the results obtained from our
measurements on $R$Fe$_2$Ge$_2$ single crystals. For all
moment-bearing rare earths (except the spherically symmetric
Gd$^{3+}$), the magnetic behavior of the crystals is extremely
anisotropic, with the moments at low temperature confined to
either the crystallographic \textit{c}-axis, or the basal plane.
This anisotropy primarily results from the CEF splitting of the
Hund's rule multiplet, whose Hamiltonian for a rare earth ion
located in tetragonal point symmetry can be written
as\cite{prat61a}

\begin{equation}
H_{CEF}= B_2^0O_2^0 + B_4^0O_4^0 + B_4^4O_4^4 + B_6^0O_6^0 +
B_6^4O_6^4, \label{eCEF}
\end{equation}

where $B_n^m$ are the Stevens coefficients related to the
geometrical arrangement of ions surrounding the rare earth, and
$O_n^m$ are the Stevens equivalent operators. If the coupling
between moments is ignored, at high enough temperatures the CEF
anisotropy for tetragonal point symmetry is governed only by the
$B_2^0$ term\cite{wang71a,bout73a}. If the isotropic inverse
susceptibility at high temperatures (without CEF splitting) is
written as $\chi^{-1}(T)=(T-\theta_p)/C$, where $C$ is the Curie
constant and $\theta_p$ is the paramagnetic Weiss temperature,
then the effect of CEF is to separate $\chi_{ab}^{-1}(T)$ and
$\chi_{c}^{-1}(T)$ by the appearance of orientation-dependent
Weiss temperatures $\theta_{ab}$ and $\theta_c$, such
that\cite{wang71a,bout73a}

\begin{equation}
(\theta_{ab}-\theta_{c})= \frac{3(2J-1)(2J+3)}{10}B_2^0,
\label{eThab-Thc}
\end{equation}

where $J$ is the total angular momentum of the Hund's rule ground
state of the rare earth. This expression shows how the magnitude
of $B_2^0$ contributes to the level of anisotropy in the
compounds, and the sign of $B_2^0$ determines whether the compound
becomes easy-axis or easy-plane. In the point charge model of CEF
we may write

\begin{equation}
B_2^0=\langle{r^2}\rangle A_2^0\alpha_J, \label{eB20}
\end{equation}

where $\langle{r^2}\rangle$ is positive by definition, and $A_2^0$
is a purely geometrical factor which can usually be considered
constant throughout a series of rare earth compounds if the
changes in lattice parameters are small. It follows that the sign
of $B_2^0$ is governed by the rare earth dependent coefficient
$\alpha_J$\cite{hut64a}, listed in table~\ref{tbl1}. Note that for
the $R$Fe$_2$Ge$_2$ series $\alpha_J<0$ results in axial moments
and $\alpha_J>0$ results in planar moments. Unfortunately we were
not able to reliably estimate the Weiss temperatures and their
trends, since these are obtained from extrapolation of the high
temperature behavior of the inverse susceptibility, and in this
series the extrapolation is highly sensitive to the non-negligible
temperature-independent susceptibility $\chi_0$ (see
table~\ref{tbl1}), which possibly indicates a large density of
states at the Fermi level $N(E_F)$. It is also likely that the
coupling between moments is still relevant at intermediate
temperatures, so a quantitative study of $B_2^0$ trends may also
require preparation of crystals for that specific purpose, such as
by diluting small fractions of a given moment-bearing rare earth
into a non-magnetic isostructural compound\cite{cho96a}, e.g.
(Y$_{1-x}R_x$)Fe$_2$Ge$_2$.

Our measurements made on all antiferromagnetically ordering
members empirically demonstrate the equivalence of $C_p(T)$ and
$d(\chi T)/dT$ for determining the transition temperatures. The
general behavior of $d\rho/dT$ was also very similar, resulting in
transition temperatures that were one or two tenths of a degree
higher at most. In table~\ref{tbl1} we list our best estimates of
the transition temperatures obtained from these three techniques.
If CEF effects are neglected, the magnetic ordering temperatures
$T_M$ across a rare earth series should be describable within the
framework of the Weiss molecular field theory, for which we may
write\cite{noak82a}

\begin{equation}
T_M=\frac{2}{3}I(g_J-1)^2J(J+1), \label{eTmdG}
\end{equation}

where $I$ is the exchange interaction parameter and
$(g_J-1)^2J(J+1)$ is the de Gennes factor (dG). The latter is
representative of the $R-R$ exchange energy, and in cases where
the magnetic interaction between the local moments of the rare
earth ions occurs indirectly via the conduction electrons (RKKY
interaction), the ordering temperatures are expected to follow a
linear dependence with dG. Figure~\ref{fTrends}b shows the Ne\'el
temperatures from table~\ref{tbl1} plotted against the dG factor.
A rough scaling with dG is found for the heavy rare earths
(Gd-Er), although significant deviation is seen especially for
$R=$Dy and Ho, indicating a relevant influence of CEF which lowers
the ordering temperatures in these compounds.

Estimation of the effective moments in the $R$Fe$_2$Ge$_2$ series
was far less sensitive to the influence of $\chi_0$ than the Weiss
temperatures. It is worth restating here that the Fe ions are
non-magnetic in this series, and the obtained values for
$\mu_{eff}$ are in good agreement with the expected moments of
free rare earth ions in their 3+ state (fig~\ref{fTrends}c),
except for Sm which resulted only about 2/3 of the expected value.
This discrepancy in SmFe$_2$Ge$_2$ is not surprising given its
distinct non-Curie-Weiss behavior at high temperatures.

The highest measured moments at 2~K also follow the general trend
expected for saturated R$^{3+}$ ions, although except for $R=$~Dy
the obtained moments are slightly lower than the expected ones.
There are several factors that may be leading to these lower
values, the most trivial one being that the highest fields
available in our magnetometers (55 and 70~kOe) are insufficient to
fully align the moments in some compounds. Other contributing
factors may include in-plane anisotropy and/or canted in-plane
moments (for the easy-plane members), singlet ground state (Tm),
or low-temperature diamagnetic contribution of the itinerant band
as described by Duong et al.\cite{duon02a}.

The heat capacity measurements were very useful in giving a basic
idea of the electronic ground states of most compounds. The
non-magnetic members presented an unusually high electronic
specific heat contribution ($\gamma\sim60$~mJ/mol~K$^2$) which
very likely persists throughout the entire series and is
consistent with the large values found for $\chi_0$. Despite the
semi-quantitative nature of the analyses leading to calculation of
the magnetic entropy (resulting from the need to estimate the
non-magnetic contributions and the behavior below our lowest
measurable temperatures), it was possible to observe the existence
of a ground state doublet for the Kramers ions Nd, Sm, Dy, Er and
an octuplet for Gd. Influence of nuclear Schottky effects
prevented a definitive quantitative analysis for the non-Kramers
ions Pr, Tb and Ho, but good fits of electronic Schottky anomalies
also helped infer the lower CEF level schemes of Dy-Tm.

All compounds that ordered above 2~K showed field-driven
metamagnetic transitions at that temperature. The transitions were
quite sharp for $R=$~Pr and Nd, less so for $R=$~Sm, Gd, Tb, and
barely noticeable as a small upward curvature for $R=$~Dy and Er.
The differences should be mostly related to the fact that for
$T=2$~K, $T/T_N\ll1$ for the first group and $\lesssim1$ for the
last. The $R$Fe$_2$Ge$_2$ series does not seem to present cases of
intermediate metamagnetic steps at 2~K, but rather a single
spin-flop transition.

The in-plane resistivity behavior of all crystals in the series
behaved very similarly, showing an ``s-shaped'' curvature up to
300~K which is different from the $R$Ni$_2$Ge$_2$ series (for
example). The similarity of all curves including those for $R=$~Y
and Lu indicates that the $R$ ion magnetism is not playing a very
significant role in the resistivity behavior above the ordering
temperatures, and the lack of any appreciable linear region
indicates that standard electron-phonon scattering is not the
dominant factor either. It is interesting to note that this
``s-shaped'' resistivity with anomalously high scattering regimes
at medium to high temperatures is frequently observed in materials
with high $\gamma$ values, such as heavy fermions. The high
temperature behavior should also be influenced by the relatively
low Debye temperatures $\Theta_D=280$~K and 240~K found for $R=$~Y
and Lu respectively. A closer look at the Fermi surfaces of
$R$Fe$_2$Ge$_2$ and how they compare to other $R$T$_2$Ge$_2$
compounds may provide a better understanding of the role of the Fe
ions in these unusual features.

\section{Conclusion}
In this work we have presented a detailed characterization of
$R$Fe$_2$Ge$_2$ single crystals grown from Sn solution. The high
quality of the crystals was attested by residual resistivities and
\textit{RRR} values in the range of 3-12 $\mu\Omega$ cm and 20-90
respectively. The crystals are also virtually free of magnetic
impurities or secondary phases, allowing the study of the
intrinsic anisotropic magnetic behavior of each compound. Strong
anisotropies arising primarily from CEF effects were observed for
all magnetic rare earths except Gd, leading to moments being
confined to either the $c$-axis or basal plane, as determined by
the sign of the Steven's coefficient $B_2^0$. Ne\'el temperatures
were determined by three independent techniques, and roughly scale
with the de Gennes factor for the heavy rare earths, although for
$R=$~Dy and Ho (and possibly Tm) the Ne\'el temperatures have been
significantly reduced by CEF effects. A second, lower temperature
transition between different magnetic phases was observed for
$R=$~Pr, Nd, Sm, and Gd. A single metamagnetic transition at 2~K
was found for all members whose moments ordered above 2~K. The
calculated effective moments per rare earth atom are close to the
expected free ion values of $R^{3+}$ except for Sm which behaves
anomalously in the paramagnetic state. Tm was the only
moment-bearing rare-earth which did not order down to 0.4~K, and
displays anomalous low temperature behavior probably due to a
non-magnetic singlet ground state, resulting in a crossover to a
Van Vleck-type susceptibility below $\sim2$~K. The non-magnetic
members of this series ($R=$~Y, Lu) showed an unusually large
electronic specific heat coefficient ($\gamma\sim60$~mJ/mol~K$^2$)
and temperature-independent susceptibility term ($\chi_0\sim0.003$
emu/mol), indicative of a relatively large density of states at
the Fermi surface. LuFe$_2$Ge$_2$ was reported for the first time
as a compound, and showed a transition at 9~K possibly due to the
formation of a charge- or spin-density wave. More detailed
investigations of individual compounds in the series, including
studies on the in-plane anisotropy, out-of-plane resistivity,
neutron diffraction, EPR, $^{57}$Fe M\"ossbauer spectroscopy, and
band structure calculations should prove useful in further
understanding their individual magnetic and electronic properties.
Substitution studies on the Fe and Ge sites may also help
understand the peculiarities of the $R$Fe$_2$Ge$_2$ compounds
compared to other members of the $R$T$_2$X$_2$ family.

We acknowledge the help of R. A. Ribeiro and C. Petrovic in the
x-ray diffraction measurements, and K. D. Myers in the early
development and characterization of the crystals. We are also
thankful to J. Schmalian for fruitful discussions. Ames Laboratory
is operated for the US Department of Energy by Iowa State
University under Contract No. W-7405-Eng-82. This work was
supported by the Director for Energy Research, Office of Basic
Energy Sciences.




\end{document}